# AI Skills Improve Job Prospects: Causal Evidence from a Hiring Experiment

Fabian Stephany[a,b,c,d], Ole Teutloff[a, e, f], Angelo Leone[a]

a) Oxford Internet Institute, University of Oxford, UK, b) Institute for New Economic Thinking, Oxford Martin School, c) Bruegel, Brussels, Belgium, d) Humboldt Institute for Internet and Society, Berlin e) Copenhagen Center for Social Data Science, University of Copenhagen, Denmark, f) Medialab, Sciences Po Paris, France

📧fabian.stephany@oii.ox.ac.uk

March 2026

**Abstract**

The growing adoption of artificial intelligence (AI) technologies has heightened interest in the labor market value of AI-related skills, yet causal evidence on their role in hiring decisions remains scarce. This study examines whether AI skills serve as a positive hiring signal and whether they can offset conventional disadvantages such as older age or lower formal education. We conducted an experimental survey with 1,725 recruiters from the United Kingdom, the United States and Germany. Using a paired conjoint design, recruiters evaluated hypothetical candidates represented by synthetically designed résumés. Across three occupations of graphic design, office assistance, and software engineering, AI skills significantly increase interview invitation probabilities by approximately 8 to 15 percentage points, compared with candidates without such skills. AI credentials, such as university or company-backed skill certificates, only lead to a moderate increase in invitation probabilities compared with self-declaration of AI skills. AI skills also partially or fully offset disadvantages related to age and lower education, with effects strongest for office assistants, for whom formal AI certificates play a significant additional compensatory role. Effects are weaker for graphic designers, consistent with more skeptical recruiter attitudes toward AI in creative work. Finally, recruiters' own background and AI usage significantly moderate these effects. Overall, the findings demonstrate that AI skills function as a powerful hiring signal and can mitigate traditional labor market disadvantages, with implications for workers' skill acquisition strategies and firms' recruitment practices.

Acknowledgments: The authors are thankful for the funding by the Microsoft AI Economy Institute. Ole Teutloff acknowledges support from the Carlsberg Foundation through an Internationalisation Fellowship grant. The authors thank the participants of the Global Technology Forum in Heilbronn (November 2025), the Microsoft AI Economy Institute seminar in Redmond (June 2025), and the Re-Skill Conference in Ferrara (November 2025) for their valuable contributions. The authors are thankful to the copy-editing of this article by David Sutcliffe.

## 1. Introduction

The demand for AI skills is growing. In major labor markets such as the US or the UK demand for AI skills has approximately tripled between the mid-2010s and the mid-2020s (Stanford HAI, 2025). As firms across industries are increasingly integrating AI systems into production, coordination, and decision-making processes, they are seeking talent skilled with AI, since unlocking AI's potential depends less on the technology itself than on the human skills needed to leverage it (Brynjolfsson et al., 2018). Consistent with basic economic theory, rising demand has translated into increasing salaries. A growing body of empirical work documents substantial wage premia associated with AI skills, with estimates suggesting that workers listing AI-related competencies command significantly higher salaries than otherwise comparable workers (Bone et al., 2025; Stephany and Teutloff, 2024).

Despite growing evidence that AI skills are associated with higher wages and improved labor market outcomes, most existing studies rely on observational data. These studies typically infer the value of AI skills from correlations in job postings, résumés, or wage data. While informative, such approaches face well-known limitations. Workers who possess AI skills may differ systematically from those who do not along unobserved dimensions, such as motivation, access to elite educational institutions (Alekseeva et al., 2021), or workforce adaptability. Indeed, AI proficiency might simply proxy for multiskilling – a trait long theorized to reduce resistance to technological change (Carmichael and MacLeod, 1993) and drive firm productivity (Kim and Park, 2003). Similarly, firms that demand AI skills may differ in ways that independently affect wages and hiring practices (Garcia-Lazaro 2025). As a result, it remains difficult to isolate whether AI skills themselves causally influence labor market outcomes or merely proxy for these other valued characteristics.

This paper provides causal evidence on the economic value of AI skills by investigating whether AI skills improve hiring prospects. We conduct a conjoint survey experiment with 1,725 professionals who have hiring experience in the UK and the US, collecting 22,195 CV comparison tasks. We present participants with pairs of fictitious CVs and ask them to choose which candidate they would invite for an interview. Each participant evaluated candidates for one of three roles: office assistant, graphic designer, or software engineer. These roles span administrative, creative, and technical work, each requiring different types of AI competencies ranging from general AI literacy to specialized technical skills. The CVs were programmatically generated to systematically vary the presence of five AI skill treatments: no AI skills mentioned, self-reported AI skills, company-certified AI skills, professional certification, and university certification. We also varied candidate characteristics including age and educational attainment to examine whether AI skills can offset traditional hiring disadvantages, such as the "technological obsolescence" stereotype often applied to older workers (Hudomiet and Willis, 2022). This experimental approach is inspired by previous literature in labor economics that use discrete choice experiments and vignette studies to elicit employer preferences (e.g., Humburg & van der Velden, 2015; Piopiunik, Schwerdt, Simon, & Woessmann, 2020).

Our experimental approach allows us to address three questions. First, do AI skills on resumes increase the probability of being invited to an interview, and does this effect differ across occupational contexts? Second, inspired by recent work on skills versus degrees (Bone et al., 2025), we investigate whether AI skills can offset traditional hiring disadvantages such as lower formal education or being older. Third, drawing on job market signaling theory (Spence, 1973) and recent work on "muddled information", that is disorganized, confusing, or ambiguous data where key details are mixed with irrelevant factors, leading to a loss of clarity. As AI might make CVs less informative of a candidate's actual ability (Frankel and Kartik, 2019), we examine whether more credible signals of AI competence carry greater weight in hiring decisions. If employers view AI skills with uncertainty, costly and verifiable credentials should provide stronger signals than self-reported claims.

Our findings show that candidates who list AI skills are substantially more likely to receive interview invitations, with the magnitude of this effect varying systematically across occupations and forms of skill disclosure. Overall, across all occupations and certification levels, AI skill signals raise interview probabilities by between 8 and 15 percentage points. Recruiters place the highest value on AI skills when hiring for technical roles: when recruiting software engineers, online-certified AI skills generate the highest interview probabilities, reaching up to 72% in our sample. For applicants without additional disadvantages, company-issued AI certificates modestly outperform self-reported skills, yielding interview probabilities of around 64%, though these incremental gains over self-reported skills are generally small, and differences across certification types are rarely statistically different from self-reported AI skills. An important exception emerges for office assistant candidates without tertiary education, for whom AI skills represented by university-issued certificates lead to the largest gains, increasing interview probabilities by up to 25 percentage points relative to comparable candidates without AI skills. This pattern indicates that while the mere presence of AI skills carries substantial signaling value in itself, formal certification can play a decisive compensatory role when conventional educational signals are absent.

To further strengthen the causal interpretation of our results, we examine how recruiters' own perceptions of and engagement with AI technologies shape their evaluation of AI skills. We find substantial heterogeneity across occupational contexts: recruiters hiring for graphic design roles express significantly greater concern about the potential harms of AI for creative work, and these concerns are closely associated with lower valuations of AI skills, helping explain the weaker effects observed for this occupation. In contrast, recruiters evaluating technical roles display far less skepticism. Moreover, we identify a pronounced gatekeeper effect linked to recruiters' own AI engagement. Recruiters who use AI tools frequently reward AI-skilled candidates much more strongly than less-engaged recruiters. Among these high-usage recruiters, interview probabilities for AI-skilled software engineering candidates approach 75% in our sample, whereas recruiters with low AI usage exhibit a markedly flattened response to AI skills.

Our study is one of the first to provide causal evidence for the growing literature on the labor market value of AI skills[1]. The diffusion of generative AI has expanded demand for workers who can develop, deploy, and effectively use AI systems (Bone et al., 2025; Stanford HAI, 2025; OECD). For instance, Teutloff et al. (2025) show how ChatGPT's release shifted demand on freelancing platforms toward AI-complementary skills, illustrating that new technologies create opportunities for workers whose capabilities complement them. This results in a substantial AI skill premium: job postings requiring AI competencies advertise about 20% higher wages than comparable positions (Bone et al., 2025). This premium reflects both scarcity and complementarity, as AI skills combine productively with diverse other skills (Stephany and Teutloff, 2024). Beyond wages, AI-related positions offer enhanced non-monetary benefits including remote work and generous leave policies (Mira et al, 2025). Only recently, studies have started to experiment with causal approaches to examine the AI skill premium (Cui et al., 2025; Firpo et al., 2025).

Our second contribution speaks directly to signaling theory. While Spence (1973) originally focused on a single dimension of ability, modern theory must account for the "muddling" of signals in the generative AI era. Frankel and Kartik (2019) propose a "Muddled Information" framework, suggesting that as signal manipulation becomes cheaper, signals become less informative about natural ability and more indicative of "gaming ability" (AI proficiency). This shift drives the labor market toward a "cheap talk" environment (Galdin and Silbert, 2025), where the ease of generating polished materials – that may appear to attribute the associated skills to their creators – makes the skills themselves a noisy signal that fails to separate high- and low-ability workers. Indeed, Cui et al. (2025) provide evidence that AI acts as a substitute for pre-existing writing ability rather than a complement, effectively narrowing the gap between workers of different initial skills. This concern is closely related to debates about the proliferation of low-quality, AI-generated content – often described as "AI slop" or "work slop" – that mimics competence without substantive effort (Niederhoffer et al., 2025). While some workers genuinely augment their capabilities through AI, others may use it as a substitute for effort, producing lower-quality output (del Rio-Chanona et al., 2025). Our study shows that, despite these concerns, AI skills function as a strong market signal even when self-reported. However, these returns are not uniform; echoing Firpo et al.'s (2025) finding that AI signal value is highly context-dependent, we observe that certification matters most for conventionally disadvantaged candidates, while skepticism toward AI is most pronounced among recruiters hiring for creative roles, where sentiment toward AI skills is markedly more negative.

[1] For an overview see Stephany & Teutloff (2026).

Our findings have important implications for managers, workers, and policymakers. For managers, the results reveal a risk of unintended hiring bias: recruiters' own engagement with AI strongly influences how they value AI skills, meaning organizations may systematically undervalue such competencies if hiring staff are not technologically engaged. Firms seeking to build AI capabilities should therefore invest in the digital literacy of recruiters, while also addressing pronounced skepticism toward AI in creative roles. Such skepticism may reflect legitimate concerns about output quality, artistic integrity, or the ethical provenance of AI-generated content, but it may equally rest on an outdated view of AI as poorly suited to creative work. For workers, particularly those facing disadvantages such as age discrimination or lacking formal degrees, AI credentials offer a viable pathway to improved employability, with compensatory effects that can partially offset structural barriers. For policymakers, the finding that AI micro-credentials can substitute for missing bachelor's degrees speaks directly to the ongoing shift toward skills-based hiring, where demonstrated competencies are gaining ground alongside formal degrees.. However, emerging adoption gaps, such as gender disparities in AI adoption and use (Humlum and Vestergaard, 2024; Stephany and Duszynski, 2026), risk translating into new forms of labor market inequality as AI skills, for example, AI-powered scheduling, reporting, content creation, software debugging and app development, become increasingly central to hiring decisions.

# 2. Survey Experiment

To causally isolate the effect of AI skills and their interactions with candidate attributes on hiring prospects, we design a paired-comparison conjoint experiment (Figure 1). This methodology allows us to estimate the causal influence of specific attributes that are often confounded in observational data (Piopiunik et al., 2020; Humburg & van der Velden, 2015). The design achieves three key objectives: it mimics the real-world comparative process recruiters use when screening candidates, it allows for clean identification of causal effects by constructing pairs that differ only on manipulated attributes, and the forced-choice format circumvents respondent-specific scale-use bias while reducing social desirability bias through revealed preference (Hainmueller et al. 2014).

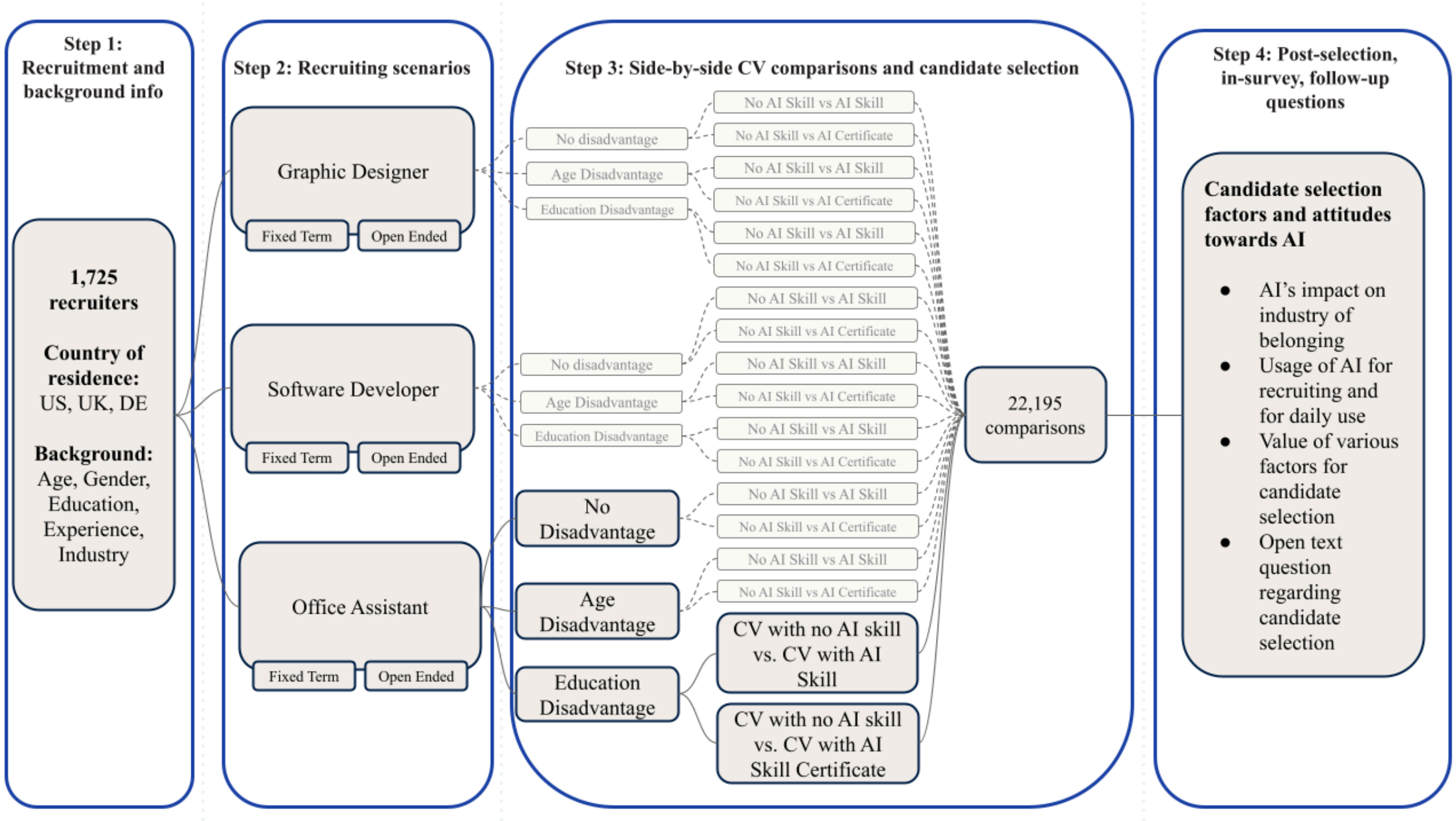

Figure 1: Structure of the conjoint experiment

### *2.1. Sample and Data Collection*

The experiment was conducted between August and September 2025 via the Prolific platform, recruiting a sample of 1,725 professionals with direct hiring experience from the United Kingdom (N=921) and the United States (N=729). Participants were screened to ensure they had been responsible for making hiring decisions in a professional capacity, including HR specialists, team leads, and line managers (see Annex F for our survey questions). This criterion ensures our results reflect the preferences of actual decision-makers rather than the general public.

Each participant completed 10 to 15 distinct choice tasks, yielding a total of 22,195 pairwise comparisons. In each task, respondents were presented with two fictitious CVs displayed side-by-side and were asked to choose which candidate they would be more likely to invite for an interview (see Annex F for example CVs). This forced-choice design mimics the comparative nature of real-world screening and reduces social desirability. Since respondents must trade off attributes across candidates rather than evaluate each in isolation, they cannot consistently favor the socially acceptable option without cost (Hainmueller et al., 2014).

### *2.2. Experimental Design*

The experimental design was structured around three key factors, which were systematically varied across the CV pairs.

**Job Profile.** The CVs were created for three distinct occupations representing a range of skill requirements and output verifiability: Office Assistant (administrative), Graphic Designer (creative), and Software Engineer (technical). For each role, we varied the contract type between a 6-month fixed-term contract and a permanent (open-ended) contract, allowing us to test whether employers' time horizons affect their valuation of AI signals.

**Candidate Attributes and Disadvantages.** Each CV was built around a core profile defined by gender, age, and education level. In a large subset of cases, one candidate (Candidate B) was presented with a potential hiring disadvantage relative to the other (Candidate A). These disadvantages were operationalized as: (a) Age: an older candidate (approximately 60 years old) paired with a younger one (approximately 32 years old); and (b) Education: a candidate with a lower level of education (e.g., an Associate's Degree) paired with a candidate with a higher level (e.g., a Bachelor's Degree).

**AI Skill Treatment.** To test our main assumptions, the disadvantaged candidate (or one of the candidates if no disadvantage occurred in the pair) was randomly assigned one of five AI skill treatments: (1) No AI skill (control), where the CV contained no mention of AI skills; (2) Self-reported AI skill, where the candidate listed AI skills in their skills section with supporting language in their job descriptions; (3) LinkedIn certification, where the candidate held an AI skill certificate from LinkedIn Learning; (4) University credential, where the candidate possessed an AI skill micro-credential from a university (e.g., "Online Certificate in Generative AI for Business Professionals, Oxford University"); or (5) Company credential, where the candidate had an AI skill certification from a major technology firm (e.g., "Generative AI for Business Professionals, IBM - Coursera certificate").

Furthermore, in a subset of cases ("disadvantage cases"), the advantaged Candidate A was assigned a relevant but non-AI professional certificate. This counter-certificate condition was included to test whether the observed effects are specific to AI credentials or attributable to certifications in general. Each role had a different counter-certification: Project Management certificate for the Office Assistant;  Advanced UX/UI certificate for the Graphic Designer; and an AWS certificate for the Software Engineer. The combinations of these characteristics were structured into 38 distinct cases (A1–A19, B1–B19), with "A cases" representing 50% of the scenarios and "disadvantage cases" the other 50%. Annex A provides a complete mapping of candidate characteristics across all experimental conditions.

### *2.3. Stimuli Construction*

To balance realism and experimental control, the fictitious CVs used as stimuli were created through a systematic, two-stage process.

**Stage 1: AI-Assisted Content Generation.** The foundational content for every CV was generated using a large language model (Google's Gemini 2.5 Pro). We first created a master design document outlining the full distribution of experimental cases, specifying the required combination of role, contract type, and candidate attributes (age, gender, education, AI treatment, and counter-certificate) for each CV. The model was then guided by a series of structured prompts to generate detailed content for each profile, including professional summaries, job histories, responsibilities, and skills. This process was iteratively refined to ensure that each CV was internally consistent, plausible, and strictly adhered to the factorial design. For example, prompts ensured that years of experience were consistent with the candidate's age and that listed skills were directly relevant to the assigned job role. The final output was a comprehensive structured dataset containing the detailed textual content for every unique CV. A crucial aspect of the design was to minimize the confounding effect of institutional prestige. Therefore, the universities from which candidates graduated were intentionally chosen to be non-elite, typically mid-tier public or state universities. This approach isolates the impact of the candidate's qualifications and our experimental treatments from the powerful brand effects of highly-ranked institutions.

**Role-Specific Profiles.** For the Office Assistant role, high education candidates held a Bachelor's degree in Business Administration, Communications, or English, while low education candidates had a high school diploma supplemented with a vocational certificate in office administration. Younger candidates were presented with 3-5 years of experience in roles such as Administrative Assistant or Team Coordinator, while older candidates had a similar number of relevant years of experience with similar titles and tasks. Analogous role-appropriate profiles were constructed for Graphic Designer and Software Engineer positions.

The AI skills were tailored to be relevant to each specific job role. For Office Assistant, AI skills focused on productivity and automation, with AI skills including: i) AI-powered scheduling, ii) Automated reporting, iii) prompt engineering for office tasks. Some skills mentioned: "Proficiency in automating workflows with Zapier and Microsoft Power Automate"; "Experience using generative AI tools for content creation and scheduling". For the Graphic Designer role, AI skills related to creative content generation, with skills including: i) Adobe AI Creative Suite, ii) AI image generation (Midjourney), ii) prompt engineering for graphic design. Some skills mentioned: "Expertise in AI-powered design tools such as Adobe Firefly and Midjourney for asset creation and ideation", "Prompt engineering for visual concept development". For the Software Engineering role, AI skills were more technical, including: i) AI-assisted debugging; ii) GenAI development, iii) prompt engineering for code, iv) Machine Learning techniques such as PyTorch. Some skills mentioned: "Skilled in integrating large language model APIs for application development".

The selection of Office Assistant, Graphic Designer, and Software Engineer is grounded in a rigorous tripartite framework of administrative, creative, and technical labor, effectively capturing the diverse impacts of the cognitive industrial revolution where all three roles exhibit 100% exposure to LLMs yet face distinct displacement or augmentation trajectories (Eloundou et al., 2024; Felten et al., 2021). For the Office Assistant, a role historically susceptible to automation-driven displacement, the specific selection of productivity-enhancing skills such as AI-powered scheduling, automated reporting, and automating workflows with Zapier is justified by the urgent need for these workers to transition from routine information processing to human-in-the-loop coordination, thereby signaling enhanced efficiency to counteract projected employment stagnation (Bureau of Labor Statistics, 2024). Conversely, the Graphic Designer represents the creative sector where the authenticity penalty creates a unique skepticism threshold regarding synthetic output; thus, skills such as Adobe AI Creative Suite, AI image generation (Midjourney), and prompt engineering for visual concept development were selected to rigorously test whether recruiters view these tools as valid augmentations of aesthetic ideation or merely as shortcuts that produce substance-less "workslop" (Edelman et al., 2024; Acemoglu, 2024). Finally, the Software Engineer serves as an example for technical synergy where AI acts as a productivity multiplier; consequently, high-complexity skills like AI-assisted debugging, GenAI development, and integrating large language model APIs were chosen to reflect the industry-wide pivot from manual coding to architectural oversight, testing the upper limits of the AI skill premium in a high-wage sector where demand for these skills has grown substantially (PwC, 2025). This strategic alignment of occupational archetypes with tailored, task-specific AI proficiencies allows the experiment to robustly interrogate signaling theory, determining how verifiable credentials function across distinct professional contexts characterised by varying degrees of information asymmetry, technological complementarity, and social valuation.

# 3. Impact of AI Skills on Hiring Decisions

## *3.1. Empirical Strategy*

To test our assumption regarding the signaling value of AI skills, we estimate two complementary sets of logistic regression models that differ in their unit of analysis and research objective.

**Model 1: Candidate-Level Analysis.** The first specification tests the main effects of AI skills and their interaction with candidate disadvantages. The unit of analysis is the individual candidate profile, and the dependent variable, $Invite_i$, is a binary indicator equal to 1 if candidate $i$ was selected and 0 otherwise. We estimate:

$$logit(P(Invite_{i,r,d} = 1)) = \alpha + \beta_1 AI_Skill_{i,r,d} + \boldsymbol{X}_{i,r,d}\gamma + \epsilon_i$$

Where *AI_Skill* is a categorical variable representing the type of AI skill treatment (No AI Skill as reference, Self-reported, LinkedIn Certificate, Company Certificate, University Certificate); *r* denotes the job role (Graphic Designer as reference, Office Assistant, Software Engineer); *d* indicates the type of disadvantage (No Disadvantage as reference, Age, Education); and ***X*** is a vector of control variables including contract type (permanent vs. 6-month) and the presence of a counter-certificate.

We estimate these models separately for candidates facing an age disadvantage (Models 1-3 in Table 1) and those facing an educational disadvantage (Models 4-6), further differentiating by job category (Software Engineer, Graphic Designer, Office Assistant) within each disadvantage group. This yields six models that allow us to examine heterogeneity in the AI skill premium across occupational contexts. Standard errors are clustered at the respondent level to account for within-recruiter correlation across the 15 choice tasks.

**Model 2: CV-Pair-Level Analysis.** To study the influence of recruiter characteristics on the valuation of AI skills, we estimate a second set of models where the unit of analysis is the CV pair rather than the individual candidate. For this analysis, we restrict the sample to pairs in which one applicant listed at least one AI skill and the other did not. The dependent variable equals 1 if the AI-skilled candidate was selected and 0 otherwise. Because our focus is on whether any AI skill provides an advantage conditional on recruiter characteristics, we aggregate all AI skill types into a single binary indicator; preliminary analyses revealed no statistically significant differences between self-reported and certificate-based skills in this specification. The model takes the form:

$$logit(P(Invite_AI_Candidate_i = 1)) = \alpha + \beta_1 AI_Skill_i + \beta_2 Recruiter_Age_i + \beta_3 Recruiter_Gender_i + \beta_4 Recruiter_Industry_i + \beta_5 Recruiter_Usage_i + \beta_6 Recruiter_Impact_i + \boldsymbol{X}_i \gamma + \epsilon_i$$

where $Recruiter_Industry$ is a dummy for recruiters from the ICT sector, $Recruiter_Usage$ is a categorical variable describing the recruiters' own usage of AI categorized as "not at all", "less than once a month", "monthly", "weekly", "daily". For Figure 5 this variable is aggregated into two categories: High (daily/weekly use) or Low (monthly/never). $Recruiter_Impact$ is a categorical variable describing how much recruiters think that AI will impact the labor market, ranging across "no impact", "slightly impactful", "moderately impactful", "very impactful", and "extremely impactful". Finally, $\boldsymbol{X}$ is a vector of control variables including role and contract type (permanent vs. 6-month).

We estimate the models in a stepwise fashion, adding blocks of variables across three stages. Step 1 (Models 1-3 in Table 2) includes job role and contract type controls only. Step 2 (Models 4-6) adds recruiter demographics, specifically recruiter age and gender. Step 3 (Models 7-9) adds the recruiter's self-reported frequency of generative AI usage. Each step includes three models corresponding to the three applicant comparison groups: no disadvantage, age disadvantage, and education disadvantage, yielding nine models in total. This stepwise approach allows us to assess the extent to which the AI skill premium is conditional on the digital literacy of the hiring gatekeeper. Standard errors in all models are clustered at the respondent level.

### *3.2. Results*

Our analysis yields robust evidence that AI skills function as a positive signal in hiring, that this signal can compensate for conventional labor market disadvantages, and that formal certification amplifies these effects in specific contexts. We structure the presentation of results around our three main assumptions: 1) AI skills increase the likelihood of a job interview invite, 2) they might compensate for conventional disadvantages related to age and education, and 3) certificates might strengthen the labor market signal of AI skills.

#### *3.2.1. AI Skills Increase the Likelihood of Getting a Job Interview*

Our data indicate that possessing any form of AI skill significantly increases the probability of receiving an interview invitation relative to having no AI skills. Figure 2 displays the predicted probabilities of interview invitation for candidates without any labor market disadvantage, pooled across job roles. The baseline for any CV comparison without disadvantaged candidates is 0.5, as each candidate has a 50-50 chance of being selected. In this first analysis, all four AI skill treatments yield predicted probabilities significantly above the 0.5 baseline that would be expected if recruiters were indifferent regarding AI skills. Even self-reported skills, which represent the lowest-cost signal in our experimental design, confer a statistically significant advantage. This finding is noteworthy because it suggests that employers do not universally discount unverified AI skill claims. Instead, the mere mention of AI proficiency appears to be valued as a signal.

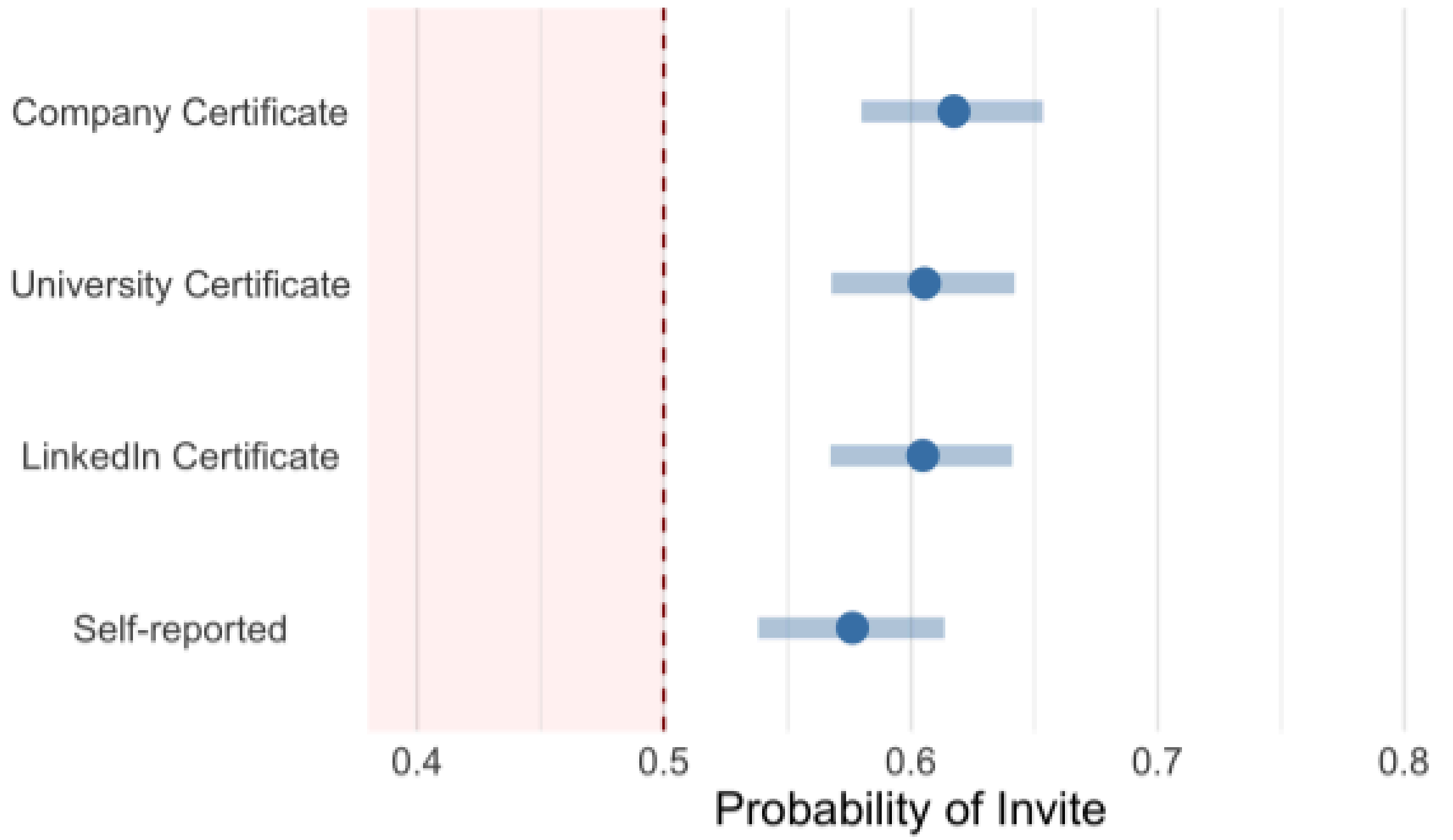

**Figure 2**: The probability of an interview invite increases significantly with the prevalence of AI skills. Certifications, such as company certificate (predicted probability = 62%), lead to a slightly higher chance of receiving an invite compared to self-reported skills. Points show predicted probabilities of receiving an interview invitation for each AI skill level, estimated from a logistic regression among non-disadvantaged applicants. Bars represent 95% confidence intervals. The dashed horizontal line at 0.5 marks the baseline of equal likelihood of invitation for either of the two applicants.

Among the certificate-based treatments, university certificates and company certificates generate the highest predicted probabilities (0.61 and 0.62, respectively), though the confidence intervals overlap substantially with those for LinkedIn certificates and self-reported skills. This suggests that, in the absence of candidate disadvantages, the marginal value of formal verification over self-reporting is modest. The primary effect is binary: having any AI skill matters more than the specific source of that skill.

Figure 3 disaggregates these results by job role, testing whether the AI skill premium differs by professional domains. For Software Engineers, all AI skill treatments generate predicted probabilities of being asked for an interview well above the 0.5 baseline, ranging from approximately 0.60 for self-reported skills to 0.71 for LinkedIn certificates. For Office Assistants, the pattern is similar though slightly attenuated with predicted probabilities ranging from approximately 0.55 to 0.6.

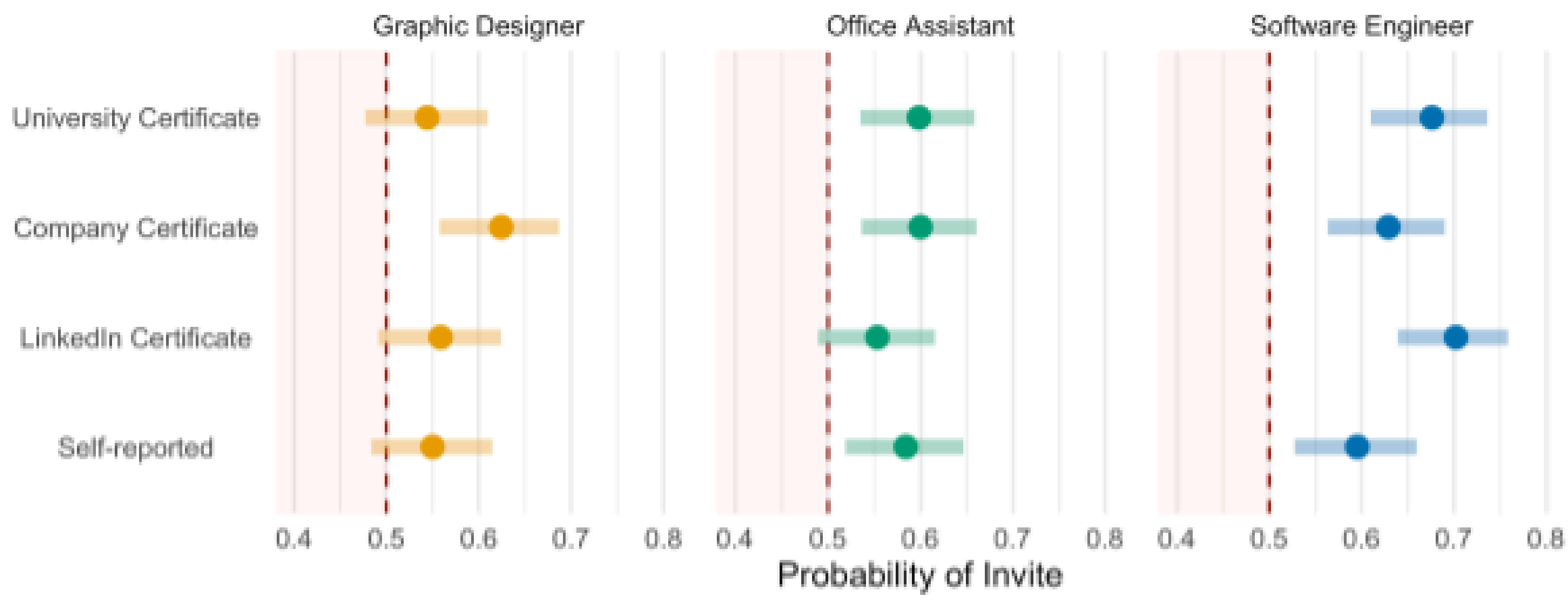

**Figure 3**: For candidates without any disadvantage, the AI skill premium is present across different occupations but varies in strength based on the professional domain. The effect is strongest for technical roles like Software Engineers, where LinkedIn certificates generate the highest interview probability (predicted probablity = 72%), whereas creative roles like Graphic Designers show notably smaller effect sizes. Points show predicted probabilities of receiving an interview invitation for each AI skill level, estimated from a logistic regression among non-disadvantaged applicants. Bars represent 95% confidence intervals. The dashed horizontal line at 0.5 marks the baseline of equal likelihood of invitation for either of the two applicants.

The effect is weaker for Graphic Designers, where self-reported AI skills hover near the 0.5 baseline and certificate-based credentials generate more modest lifts (predicted probabilities of approximately 0.55-0.62). This weaker effect for creative roles is consistent with an authenticity penalty interpretation: recruiters may view AI-generated design outputs with skepticism, questioning whether candidates who rely on tools like Midjourney or Adobe Firefly possess genuine creative ability. We find corroborating evidence in our sentiment analysis of recruiter open-text responses (for details on the methodological approach see Annex C). Among recruiters evaluating Graphic Designer positions, 43.8% expressed negative sentiment toward AI skills compared to only 49.0% positive. This is nearly an even split, which contrasts sharply with recruiters for Office Assistants (24.8% negative, 70.1% positive) and Software Engineers (22.0% negative, 72.9% positive). The skepticism toward AI among Graphic Design recruiters aligns with the lower returns for AI skills and certificates we observe for this occupation.

### *3.2.2. AI Skills Can Compensate for Conventional Hiring Disadvantages*

We assume that AI skills compensate for conventional hiring disadvantages faced by older applicants and applicants with lower levels of formal education. Figure 4 displays the predicted probabilities of interview invitation for candidates facing either an age disadvantage (Panel A) or an education disadvantage (Panel B), disaggregated by job role and AI skill treatment. Table 1 presents the corresponding regression coefficients.

Table 1 shows the results of a set of logistic regression models estimating the probability that an applicant is invited to an interview (1 = invited; 0 = not invited). The main explanatory variable is the type of AI skill listed by the applicant. The reference category is applicants without any AI skills, and we compare this group to applicants reporting one of four AI skill types: self-reported AI skills, LinkedIn certificates, company-issued certificates, and university-issued certificates. We estimate these models separately for two groups of applicants: (1) those facing an age disadvantage (Columns 1–3), and (2) those facing an educational disadvantage (columns 4–6). Within each disadvantage group, we additionally differentiate by job category: Software Engineer, Graphic Designer, and Office Assistant, producing six models in total. The sample sizes of the individual models vary slightly due to missing values.

Across almost all specifications, the coefficients indicate that possessing any type of AI skill increases the odds of being invited to an interview relative to having no AI skills. However, the magnitude of the effect varies both by job type and by the kind of disadvantage faced by applicants.

**Age Disadvantage[2].** For applicants with an age disadvantage (approximately 60 years old compared to 32 years old), the positive effects of AI skills are consistently present, though their strength differs across occupations. Figure 4A shows that older candidates without AI skills face a substantial hiring penalty: predicted probabilities of interview fall to approximately 0.30 for Graphic Designers, 0.35 for Office Assistants, and 0.38 for Software Engineers. They are all well below the 0.5 baseline and indicate a strong preference for younger competitors. These baseline penalties are consistent with the technological obsolescence stereotype often attached to older workers (Hudomiet & Willis, 2023). In Table 1 (Models 1-3), the coefficient for self-reported skills in the Software Engineers specification ($\beta = 0.97, p < 0.01$) is the largest in magnitude across all conditions. The lowest coefficient is reported for LinkedIn certificates among applicants in Graphic Design ($\beta = 0.29, p < 0.1$). Beyond technological concerns, these penalties likely also reflect employer perceptions regarding the limited future career potential and shorter investment horizons associated with older applicants (Posthuma & Campion, 2009; North & Fiske, 2012). However, the addition of AI skills substantially recovers these probabilities. For Software Engineers with an age disadvantage, company and university certificates lift predicted probabilities to approximately 0.44–0.5, effectively neutralizing the age penalty. For Office Assistants, older candidates with verified AI credentials achieve predicted probabilities near or above 0.50. The smallest effects are observed among Graphic Designers, where AI skills appear to matter least for interview selection, consistent with the negative sentiment patterns documented above.

**Education Disadvantage.** For applicants with an educational disadvantage (e.g., an associate's degree rather than a bachelor's degree), we observe a similar general pattern, but with notable differences in the Office Assistant role. Figure 4B and the corresponding coefficients in Table 1 (Models 4-6) show that certification-based skills, and especially university-issued certificates, generate the strongest positive effect on interview chances, clearly outweighing the impact of self-reported skills. The coefficient for university certification in the Office Assistant specification ($\beta = 1.02, p < 0.01$) is the largest in magnitude across all conditions and significantly exceeds the coefficient for self-reported skills ($\beta = 0.93, p < 0.01$). This suggests that for lower-skilled or administrative roles, formal validation of AI competencies may matter more than informal or self-declared skills. University micro-credentials appear to function as a partial substitute for a missing bachelor's degree.

[2] We also tested a gender disadvantage operationalized as female applicants with and without AI skills competing with male applicants without AI skills. Figure A2 shows the results which indicate that female applicants are more likely to be invited to an interview than their male counterparts, regardless of level of AI skills. Gender does not seem to mediate the role of AI skills.

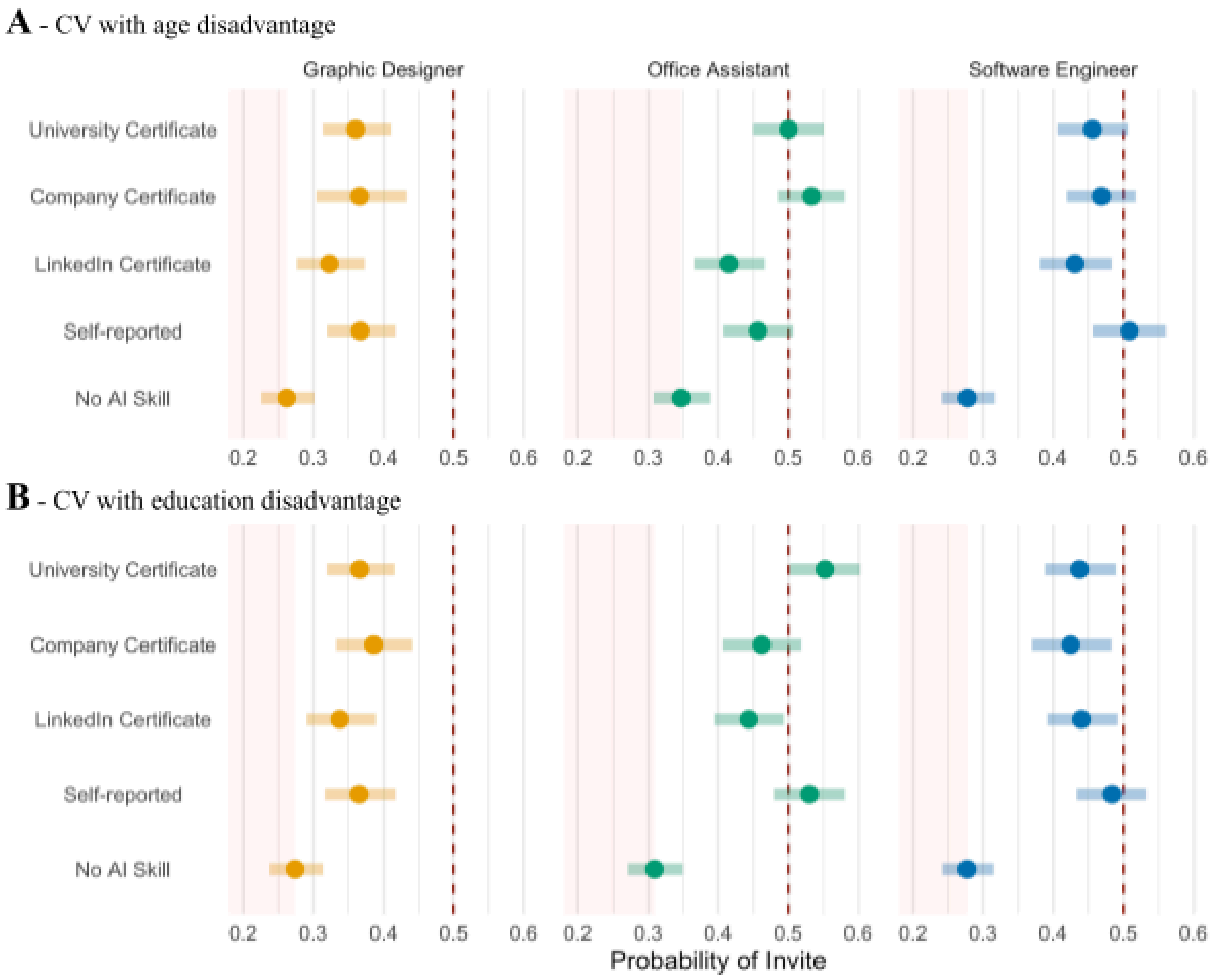

**Figure 4**: AI skills act as a powerful compensatory signal, significantly offsetting the hiring penalties associated with older age (Panel A) and lower formal education (Panel B). For age-disadvantaged Software Engineers, verified certifications effectively neutralize the age penalty (predicted probablity = 50-55%), while university micro-credentials provide the strongest boost for administrative candidates lacking a bachelor's degree, acting as a partial substitute for traditional qualifications. Points show predicted probabilities of receiving an interview invitation for each AI skill level, estimated from a logistic regression. Bars represent 95% confidence intervals. The shaded red area indicates the baseline probability of being invited for an interview for CVs without AI skills. The dashed horizontal line at 0.5 marks equal likelihood of invitation for either of the two applicants.

Overall, the results strongly support the hypothesis that AI skills generally improve applicants' prospects of being invited to an interview even when those applicants face conventional hiring disadvantages, though the type of AI skill, the job context, and the specific disadvantage condition the magnitude of this compensatory effect.

.

**Table 1.** AI skills and invitation probability across occupations. Compared to candidates without AI skills, those with AI skills show consistently higher invitation probabilities across all models. Effects are estimated separately for software engineers, graphic designers, and office assistants, accounting for age disadvantage (Models 1–3) and educational disadvantage (Models 4–6)

Table 1: Individual Level Logistic Regression

| | *Dependent variable: Candidate gets invited (0/1)* | | | | | |
|---|---|---|---|---|---|---|
| | Age disadvantage | | | Educational disadvantage | | |
| | Software Engineer | Graphic Designer | Office Assistant | Software Engineer | Graphic Designer | Office Assistant |
| **AI Skills (ref. No AI skills)** | | | | | | |
| Self-reported | 0.97*** | 0.45*** | 0.49*** | 0.86*** | 0.40*** | 0.93*** |
| | (0.14) | (0.14) | (0.14) | (0.14) | (0.14) | (0.14) |
| LinkedIn certificate | 0.60*** | 0.29* | 0.30** | 0.67*** | 0.28* | 0.59*** |
| | (0.14) | (0.15) | (0.14) | (0.14) | (0.15) | (0.14) |
| Company certificate | 0.74*** | 0.47*** | 0.76*** | 0.64*** | 0.51*** | 0.66*** |
| | (0.14) | (0.17) | (0.13) | (0.15) | (0.15) | (0.15) |
| University certificate | 0.71*** | 0.46*** | 0.65*** | 0.71*** | 0.41*** | 1.02*** |
| | (0.14) | (0.14) | (0.14) | (0.14) | (0.14) | (0.14) |
| Constant | −1.24*** | −0.97*** | −0.79*** | −1.26*** | −0.60** | −0.10 |
| | (0.26) | (0.27) | (0.26) | (0.26) | (0.26) | (0.29) |
| Observations | 2,102 | 1,921 | 2,133 | 2,085 | 2,010 | 2,030 |
| Log Likelihood | −1,394.69 | −1,188.93 | −1,439.23 | −1,359.58 | −1,273.54 | −1,353.08 |
| Akaike Inf. Crit. | 2,807.39 | 2,395.86 | 2,896.46 | 2,737.17 | 2,565.08 | 2,724.15 |
| *Note:* | | | | | | *p<0.1; **p<0.05; ***p<0.01 |

### *3.2.3. Certifications Modestly Increase Employability in the Presence of Candidate Disadvantages*

Our last assumption posits that formal certification, represented by micro-credentials from LinkedIn, companies, or universities, further increases employability beyond self-reported AI skills. The evidence for the additional value of credentials is nuanced and context-dependent.

As already shown, in the absence of candidate disadvantages, the marginal value of formal verification over self-reporting is modest (Figures 2 and 3). All four AI skill types generate predicted probabilities that are statistically distinguishable from 0.5, but the confidence intervals overlap across certification levels. This suggests that when candidates have no obvious deficits, the primary effect is binary: having any AI skill matters more than the specific source of that skill.

However, our assumption about the usefulness of credentials receives strong support in the context of candidate disadvantages, particularly for education-disadvantaged applicants in administrative roles. As documented above (Figure 4B), university-issued certificates show the strongest positive effect on interview chances for Office Assistants with lower educational attainment, clearly outweighing the impact of self-reported skills.

### *3.2.4. Recruiters' Usage and Perception of AI Influence the Value of AI Skills*

While several recruiter characteristics influence AI skill valuation, two patterns stand out. Firstly, recruiters who frequently use AI themselves perceive candidate AI skills very differently than those who rarely use it. Secondly, recruiters' perception of the value brought by AI skills varies based on occupations. Table 2 presents the CV-pair-level regressions, and Figure 5 visualizes the split-sample results for high-AI-usage recruiters (Panel A: daily or weekly users) versus low-AI-usage recruiters (Panel B: monthly use or less).

Table 2 presents logistic regression results where the unit of analysis is the CV pair. In each pair, one applicant listed at least one AI skill and the other did not. The dependent variable equals 1 if the candidate with any AI skill in the pair was invited to interview, and 0 otherwise. Because our focus is on whether *any* AI skill provides an advantage, we restrict the analytical sample to CV pairs in which one applicant reported an AI skill. We do not distinguish between different types of AI skills in these models, as preliminary analyses showed no statistically significant differences between self-reported and certificate-based skills. We estimate the models in a stepwise fashion, adding blocks of variables across five distinct specifications to isolate the drivers of recruiter preference:

- Baseline: This model controls solely for the randomized experimental design features, specifically the Role Type and Contract Type.
- \+ Recruiter Characteristics: This specification extends the baseline by adding recruiter demographics, including age, gender (male), and whether the recruiter is recruiting for the ICT sector.
- \+ Recruiter AI Usage: Building on the characteristics model, this column adds the recruiter's self-reported frequency of generative AI use, categorized into four levels ranging from "Less than once a month" to "Daily".
- \+ Recruiter AI Impact: This alternative specification replaces the usage variables with the recruiter's belief regarding the AI impact on the industry they are primarily hiring for, measured on an ordinal scale from "Slightly impactful" to "Extremely impactful," while maintaining demographic controls.
- Full Model: The comprehensive model simultaneously includes all covariates – role, contract, recruiter demographics, AI usage frequency, and perceived AI impact – to determine which factors remain significant when controlling for the others.

**Table 2.** Invitation probability of AI-skilled candidates in paired CV evaluations. Using CV pairs as the unit of observation, the dependent variable indicates whether the AI candidate is invited. All models control for job and recruiter characteristics. Recruiters working for ICT firms, with higher usage of AI and a stronger believe that AI will change the labour market are more likely to pick the AI-skilled candidate.

| | *Dependent variable: AI candidate gets invite (0/1) - No Disadvantage* | | | | |
|---|---|---|---|---|---|
| | Baseline | + Recruiter Characteristics | + Recruiter AI Usage | + Recruiter AI Impact | Full Model |
| **Controls** | | | | | |
| Role Type | YES | YES | YES | YES | YES |
| Contract Type | YES | YES | YES | YES | YES |
| **Recruiter** | | | | | |
| Age | | 0.02*** | 0.02*** | 0.02*** | 0.02*** |
| | | (0.004) | (0.004) | (0.004) | (0.004) |
| Male (ref. Female) | | 0.13 | 0.06 | 0.09 | 0.06 |
| | | (0.08) | (0.08) | (0.08) | (0.08) |
| **Recruiting for ICT (ref. Other)** | | 1.06*** | 0.92*** | 0.92*** | 0.87*** |
| | | (0.23) | (0.23) | (0.23) | (0.23) |
| **AI Usage (ref. Never)** | | | | | |
| Less than once a month | | | 0.57*** | | 0.42** |
| | | | (0.18) | | (0.19) |
| Monthly | | | 1.07*** | | 0.83*** |
| | | | (0.19) | | (0.20) |
| Weekly | | | 1.00*** | | 0.67*** |
| | | | (0.17) | | (0.18) |
| Daily | | | 1.20*** | | 0.75*** |
| | | | (0.17) | | (0.18) |
| **AI impact on labour market (ref. No impact)** | | | | | |
| Slightly impactful | | | | 0.57*** | 0.46** |
| | | | | (0.21) | (0.22) |
| Moderately impactful | | | | 0.98*** | 0.77*** |
| | | | | (0.21) | (0.21) |
| Very impactful | | | | 1.32*** | 1.05*** |
| | | | | (0.21) | (0.22) |
| Extremely impactful | | | | 1.54*** | 1.25*** |
| | | | | (0.23) | (0.24) |
| Constant | 0.24*** | −0.51*** | −1.37*** | −1.44*** | −1.83*** |
| | (0.08) | (0.19) | (0.24) | (0.27) | (0.29) |
| Observations | 2,765 | 2,659 | 2,659 | 2,659 | 2,659 |
| Log Likelihood | −1,849.64 | −1,750.86 | −1,718.05 | −1,707.09 | −1,695.55 |
| Akaike Inf. Crit. | 3,707.27 | 3,519.71 | 3,462.09 | 3,440.19 | 3,425.09 |
| *Note:* | | | | | *p<0.1; **p<0.05; ***p<0.01 |

**Recruiter demographics and sector.** Recruiter characteristics play a significant role in hiring preferences. Older recruiters are significantly more likely to select the AI-skilled candidate ( $\beta = 0.02$, $p < 0.01$), suggesting that seniority does not dampen openness to AI-related qualifications. While the direction for male recruiters is positive, the effect is not statistically significant in the full model ($\beta = 0.06$), as male recruiters tend to use AI more frequently (see Table A1 in the Appendix). Notably, recruiters hiring for the ICT sector show a massive preference for AI skills compared to other sectors, with a highly significant coefficient of $\beta = 1.06$ ($p < 0.01$) in model 2, settling at ($\beta = 0.87$, $p < 0.01$) in the full model.

**Recruiter's AI usage.** A robust pattern emerges when we incorporate the recruiter's own generative AI usage. Frequency of AI use is strongly and positively associated with selecting the AI-skilled candidate. This effect follows a clear monotonic gradient: recruiters who use generative AI more frequently are progressively more likely to pick the AI candidate. In the usage-specific model (column 3 of table 2), the coefficients increase from $0.57$ ($p < 0.01$) for "less than once a month" users to $1.20$ ($p < 0.01$) for "daily" users. Even in the full model, where AI impact beliefs are controlled for, daily users maintain a strong preference ($\beta = 0.75$, $p < 0.01$).

**Recruiter's perception of AI impact.** Table 2 reveals that beyond usage, the recruiter's belief in the technology's impact on the industry they recruit for is a decisive factor. Recruiters who believe AI will be "Extremely impactful" on their industry show the strongest propensity to invite AI-skilled candidates, with a coefficient of $\beta = 1.54$ ($p < 0.01$) in column 4. This effect dominates the model; in the full model, the coefficient for "Extremely impactful" ($\beta = 1.25$) is the strongest predictor among all recruiter traits. This suggests that the "gatekeeper effect" is driven not just by familiarity (usage), but by a conviction that AI represents a fundamental shift in the industry.

Overall, Table 2 demonstrates that the AI skill premium is partly conditional on the recruiter's profile: older recruiters, those in the ICT sector, frequent AI users, and – most significantly – those who believe AI will drastically reshape their industry, are the primary drivers of demand for AI skills. Firms whose recruiters underestimate AI's impact or lack personal experience with the tools may be systematically undervaluing these competencies.

This pattern appears clearly in Figure 5. Among recruiters with high AI usage (daily or weekly), AI-skilled Software Engineers achieve predicted probabilities approaching 0.75, and even AI-skilled Graphic Designers exceed 0.55. Among recruiters with low AI usage (monthly or never), the entire distribution shifts downward and flattens: predicted probabilities cluster near 0.50–0.55 across all skill types and roles, and many confidence intervals include or approach 0.50. The gap between high and low AI usage recruiters appears most clearly for Office Assistants, where all credential types show substantial separation.

This finding has profound implications for labor market efficiency. It suggests that the AI skill premium documented in this study might be conditional on the digital literacy of the hiring gatekeeper. Firms whose recruiters do not themselves use generative AI may be systematically undervaluing AI competencies in candidates, creating a friction between the nominal demand for AI skills and the actual ability of hiring processes to reward them.

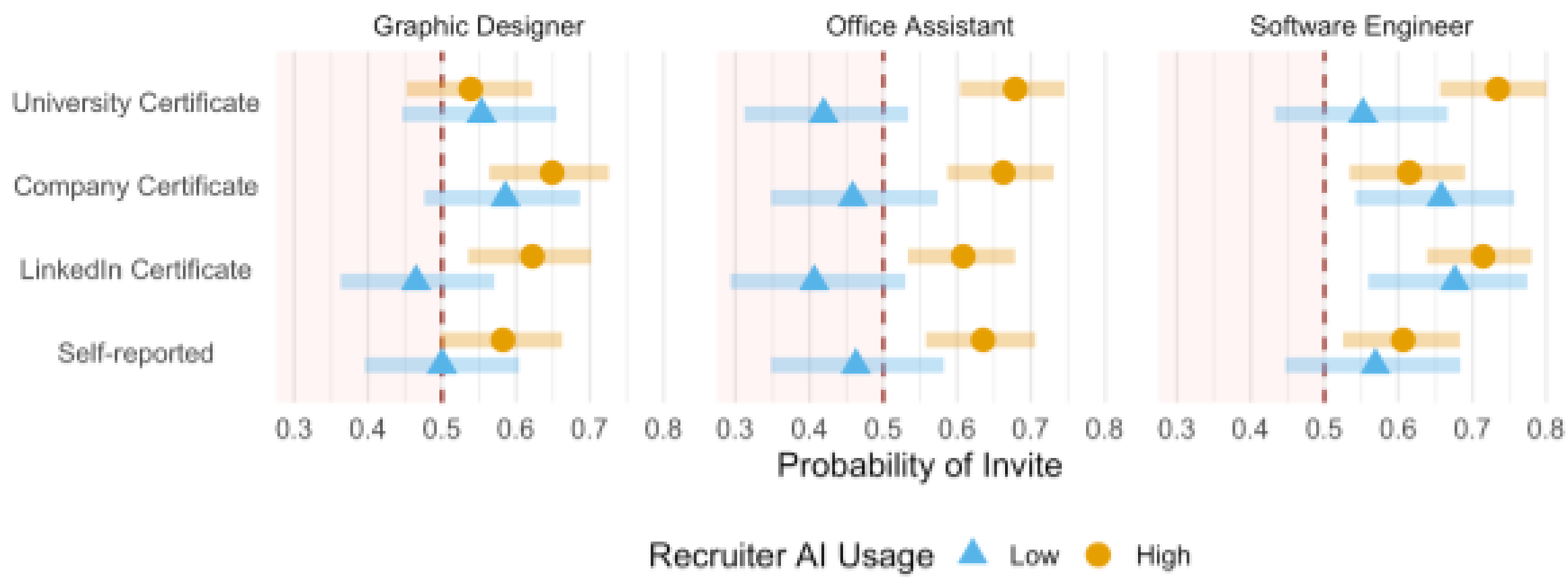

**Figure 5**: The valuation of AI skills is heavily dependent on the recruiter's own technological engagement, revealing a pronounced gatekeeper effect. Recruiters with high AI usage (daily or weekly) strongly reward AI-skilled candidates, approaching a 75% interview probability for Software Engineers, whereas recruiters with low AI usage (monthly or less) show a flattened response, appearing less able to distinguish or value AI competence. Points show predicted probabilities of receiving an interview invitation for each AI skill level, estimated from a logistic regression. Bars represent 95% confidence intervals. The dashed horizontal line at 0.5 marks the baseline of equal likelihood of invitation for either of the two applicants.

Beyond the binary metric of adoption, the perception of AI's utility – captured through the sentiment analysis detailed in Annex C – offers a critical explanatory layer for the observed premium. While recruiter usage of AI tools is positively correlated with the likelihood of selecting AI-skilled candidates, the qualitative evidence suggests this relationship is driven by an underlying belief in the technology's value proposition. As shown in Table C1, recruiters who expressed positive sentiment toward AI were significantly more likely to prioritize AI literacy, with a positive interaction effect of 5.2 percentage points (Model 2). Crucially, this "enthusiasm effect" remains robust (4.8 percentage points, Model 3) even when controlling for the recruiter's own frequency of AI usage, indicating that the premium is not merely a function of technical familiarity, but is shaped by the specific value they ascribe to the technology.

This perception of value is not uniform, however; it is mediated by the specific demands of the occupation (Table C2). The sentiment-driven premium is most pronounced for administrative roles, where positive recruiter sentiment increases the probability of selecting an AI-skilled Office Assistant for an interview by 7.1 percentage points (Model 2). The effect is lower for Software Engineers (3.5 percentage points) and Graphic Designers (4.1 percentage points). This heterogeneity implies that for administrative tasks, AI is perceived as an unambiguous efficiency multiplier, automating routine friction. Conversely, for code-related and creative occupations, the lower translation of sentiment into interview preference suggests a more complex valuation, where AI is seen as a tool requiring careful oversight rather than a direct substitute for human expertise.

## 4. Concluding remarks

This study bridges the gap between the surging demand for AI skills and the scarcity of causal evidence regarding their value in hiring. By deploying a paired conjoint experiment with 1,725 recruiters in the UK and US, we isolate the causal premium of AI credentials from confounding factors typical of observational data. Our findings confirm that AI skills act as a powerful signal of employability, increasing the probability of receiving an interview invitation by approximately 8 to 15 percentage points across diverse occupations. This effect persists regardless of whether the skills are self-taught or formally certified, suggesting that in the current labor market, the demonstrated ability to use Generative AI is valued as a proxy for adaptability and potential productivity, independent of the prestige of the credentialing institution.

Beyond the aggregate premium, our results highlight the potential of AI upskilling to act as an equalizer in the labor market. We document a significant substitution effect where AI credentials partially or fully offset the penalties associated with lower formal education and older age. For older candidates, possessing AI skills effectively neutralizes ageism, signaling that the applicant is technically current and adaptable. Similarly, for candidates without university degrees, AI proficiency serves as a compensatory mechanism, allowing them to compete with university-educated peers. This suggests that AI adoption could facilitate social mobility and extend the economic relevance of the workforce in an aging society, provided that access to training is equitable.

However, the valuation of these skills is heavily contingent on the "gatekeeper effect" we identified: a recruiter’s own familiarity with AI dictates how they value a candidate’s AI skills. Recruiters who utilize Generative AI in their daily workflows assign a significantly higher premium to these skills compared to non-users. This heterogeneity implies that the labor market integration of AI is currently asymmetric; as AI tools become more ubiquitous among hiring managers, the penalty for lacking these skills may increase, potentially deepening the divide between the AI-literate and the AI-illiterate workforce.

Theoretically, our findings contribute to the economics of education and signaling theory by distinguishing between the signaling value of traditional degrees and emerging technical competencies. While traditional education signals long-term persistence and general cognitive ability, AI skills appear to signal immediate readiness for technological disruption. This distinction is crucial for understanding how labor markets adjust to rapid technological change. We provide empirical support that targeted micro-credentialing offers a valued response mechanism that could compensate for disadvantages in formal long-term degrees as well as age penalties.

For policymakers and practitioners, the implications are two-fold. For firms, the strong preference for AI-skilled candidates – driven largely by AI-literate recruiters – suggests that organizations need to standardize how they evaluate these competencies to avoid idiosyncratic hiring biases. For policymakers, the evidence that AI skills can mitigate disadvantages for older and less-educated workers provides a compelling case for public investment in AI literacy programs. Supporting mid-career retraining in Generative AI could be a cost-effective strategy to enhance labor force participation and reduce structural unemployment among vulnerable demographic groups.

Finally, while our experimental design allows for causal identification, it is not without limitations. First, our results rely on stated preferences in a hypothetical hiring scenario; while conjoint experiments are proven predictors of real-world behavior, actual hiring decisions may involve additional constraints. Second, our analysis is limited to three specific white-collar occupations, and the magnitude of the AI premium likely varies across other sectors. Future research should examine whether the hiring premium translates into wage premiums and actual on-the-job productivity gains, as well as how the signaling value of AI skills evolves as these technologies move from novel differentiators to standard job requirements.

## Annex A - Breakdown of pairwise CV comparisons

| **Index** | **CV 1** | **CV 2** | **Disadvantage for CV2** | **AI Treatment to CV 2** | **Gender difference** |
|---|---|---|---|---|---|
| A1 | Female + Young + High Edu | Female + Young + High Edu | No disadvantage | AI: Self-reported | Same gender |
| A2 | Female + Young + High Edu | Female + Young + High Edu | No disadvantage | AI: University credential | Same gender |
| A3 | Female + Young + High Edu | Female + Young + High Edu | No disadvantage | AI: Company credential | Same gender |
| A4 | Female + Young + High Edu | Female + Young + High Edu | No disadvantage | AI: LinkedIn | Same gender |
| A5 | Male/Female + Young + High Edu | Male/Female + Young + **Low Edu** | Low Edu | None | Same gender |
| A6 | Male/Female + Young + High Edu | Male/Female + Young + **Low Edu** | Low Edu | AI: Self-reported | Same gender |
| A7 | Male/Female + Young + High Edu | Male/Female + Young + **Low Edu** | Low Edu | AI: University credential | Same gender |
| A8 | Male/Female + Young + High Edu | Male/Female + Young + **Low Edu** | Low Edu | AI: Company credential | Same gender |
| A9 | Male/Female + Young + High Edu | Male/Female + Young + **Low Edu** | Low Edu | AI: LinkedIn | Same gender |
| A10 | Male/Female + Young + High Edu | Male/Female + **Old** + High Edu | Older | None | Same gender |
| A11 | Male/Female + Young + High Edu | Male/Female + **Old** + High Edu | Older | AI: Self-reported | Same gender |
| A12 | Male/Female + Young + High Edu | Male/Female + **Old** + High Edu | Older | AI: University credential | Same gender |
| A13 | Male/Female + Young + High Edu | Male/Female + **Old** + High Edu | Older | AI: Company credential | Same gender |
| A14 | Male/Female + Young + High Edu | Male/Female + **Old** + High Edu | Older | AI: LinkedIn | Same gender |
| A15 | Male + Young + High Edu | **Female** + Young + High Edu | Female | None | Different gender |
| A16 | Male + Young + High Edu | **Female** + Young + High Edu | Female | AI: Self-reported | Different gender |
| A17 | Male + Young + High Edu | **Female** + Young + High Edu | Female | AI: University credential | Different gender |
| A18 | Male + Young + High Edu | **Female** + Young + High Edu | Female | AI: Company credential | Different gender |
| A19 | Male + Young + High Edu | **Female** + Young + High Edu | Female | AI: LinkedIn | Different gender |
| A20 | Male/Female + Young + High Edu | Male/Female + Young + High Edu | None | None | Same gender |
| B1 | Male + Young + High Edu | Male + Young + High Edu | No disadvantage | AI: Self-reported | Same gender |
| B2 | Male + Young + High Edu | Male + Young + High Edu | No disadvantage | AI: University credential | Same gender |
| B3 | Male + Young + High Edu | Male + Young + High Edu | No disadvantage | AI: Company credential | Same gender |
| B4 | Male + Young + High Edu | Male + Young + High Edu | No disadvantage | AI: LinkedIn | Same gender |
| B5 | Male/Female + Young + High Edu | Male/Female + Young + **Low Edu** | Low Edu | None | Same gender |
| B6 | Male/Female + Young + High Edu + Counter Certificate | Male/Female + Young + **Low Edu** | Low Edu | AI: Self-reported | Same gender |
| B7 | Male/Female + Young + High Edu + Counter Certificate | Male/Female + Young + **Low Edu** | Low Edu | AI: University credential | Same gender |
| B8 | Male/Female + Young + High Edu + Counter Certificate | Male/Female + Young + **Low Edu** | Low Edu | AI: Company credential | Same gender |
| B9 | Male/Female + Young + High Edu + Counter Certificate | Male/Female + Young + **Low Edu** | Low Edu | AI: LinkedIn | Same gender |
| B10 | Male/Female + Young + High Edu | Male/Female + **Old** + High Edu | Older | None | Same gender |
| B11 | Male/Female + Young + High Edu + | Male/Female + **Old** + High Edu | Older | AI: Self-reported | Same gender |
| | Counter Certificate | | | | |
| B12 | Male/Female + Young + High Edu + Counter Certificate | Male/Female + **Old** + High Edu | Older | AI: University credential | Same gender |
| B13 | Male/Female + Young + High Edu + Counter Certificate | Male/Female + **Old** + High Edu | Older | AI: Company credential | Same gender |
| B14 | Male/Female + Young + High Edu + Counter Certificate | Male/Female + **Old** + High Edu | Older | AI: LinkedIn | Same gender |
| B15 | Male + Young + High Edu | **Female** + Young + High Edu | Female | None | Different gender |
| B16 | Male + Young + High Edu + Counter Certificate | **Female** + Young + High Edu | Female | AI: Self-reported | Different gender |
| B17 | Male + Young + High Edu + Counter Certificate | **Female** + Young + High Edu | Female | AI: University credential | Different gender |
| B18 | Male + Young + High Edu + Counter Certificate | **Female** + Young + High Edu | Female | AI: Company credential | Different gender |
| B19 | Male + Young + High Edu + Counter Certificate | **Female** + Young + High Edu | Female | AI: LinkedIn | Different gender |

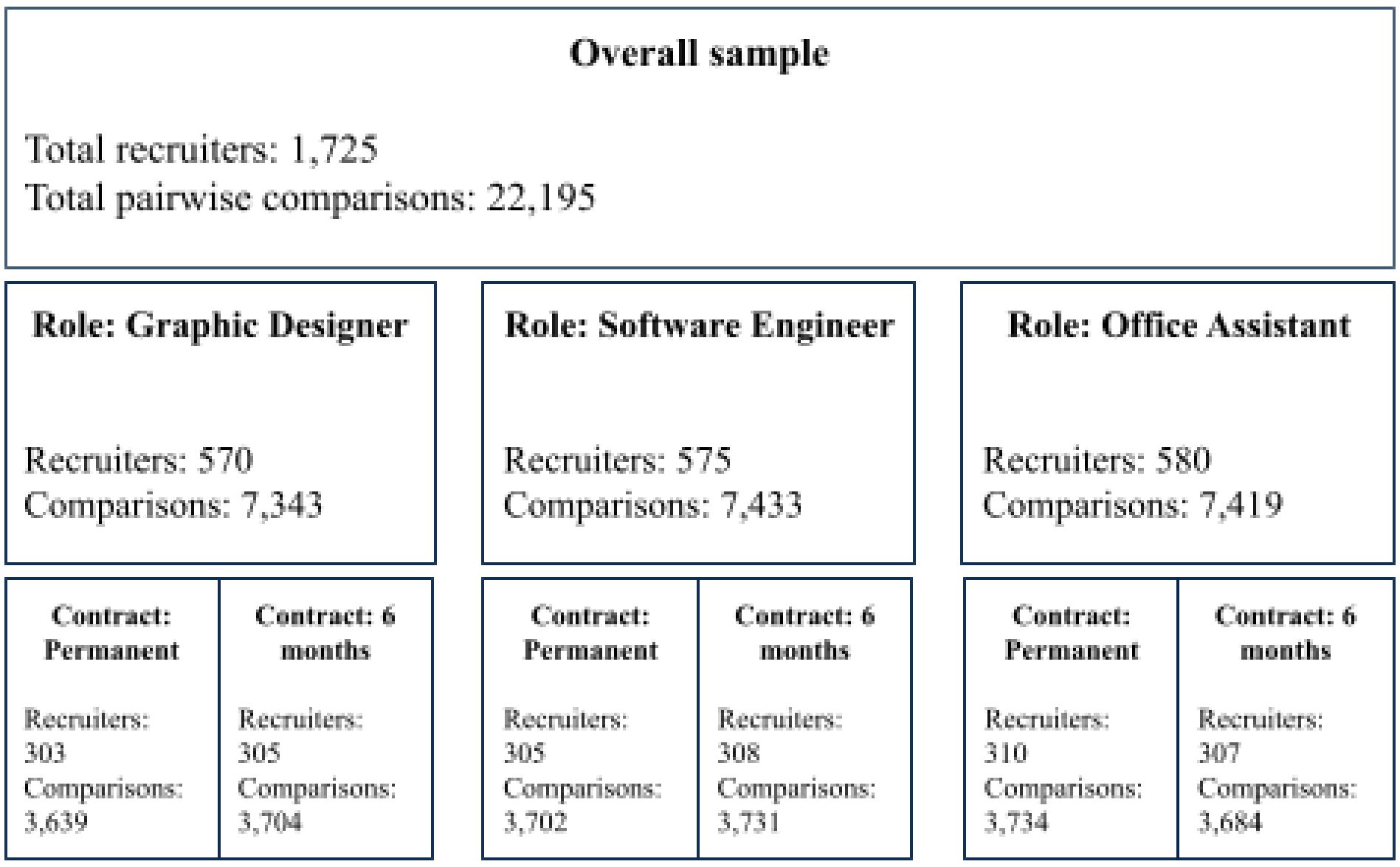

## Annex B - Job descriptions

This section details the job descriptions shown to the recruiters before proceeding to the choice tasks.

### Office assistant, 6-month contract

You will now be shown pairs of candidate profiles for the six months role of Office Assistant at an undefined organization and industry. For each pair, please carefully review both candidate profiles. Then, select the candidate (Candidate A or Candidate B) whom you would be more likely to invite for an interview for that specific role. There are no 'right' or 'wrong' answers; we are interested in your professional judgment. Please make your choices as if you were genuinely hiring for this role.

Please find below the job description for the role of Office Assistant:

Position: Office Assistant

Job Type: Full-time, 6-month contract

Overview: We are seeking a reliable, proactive and organized Office Assistant to drive efficiency in our day-to-day administrative operations. This role requires exceptional attention to detail, excellent communication skills, and the ability to manage multiple tasks with foresight.

Key Responsibilities:

- Manage calendars, proactively schedule meetings to optimize time for all participants, and coordinate appointments.
- Handle incoming communications (phone, email, mail), employing systems to prioritize and pre-draft responses.
- Maintain organized filing systems (physical and digital) designed for rapid information retrieval.
- Prepare documents, synthesize information for reports, and draft professional correspondence.
- Assist with data entry and ensure data integrity through validation and cleanup.

- Support office supply management and implement predictive inventory tracking.
- Provide administrative support to team members.
- Coordinate with vendors and service providers as needed.

Required Qualifications:

- High school diploma or equivalent 2+ years of administrative or clerical experience.
- Proficiency in Microsoft Office Suite (Word, Excel, Outlook).
- Strong written and verbal communication skills.
- Excellent organizational and time management abilities.
- Professional demeanor and customer service orientation.
- Ability to work independently and as part of a team.

Preferred Qualifications:

- Associate's degree or higher in Business Administration or related field.
- Knowledge of basic bookkeeping or accounting principles.
- Experience with modern productivity and office management software that enhances efficiency.
- Familiarity with basic data analysis or business intelligence concepts.

### Office assistant, permanent contract

You will now be shown pairs of candidate profiles for the permanent role of Office Assistant at an undefined organization and industry. For each pair, please carefully review both candidate profiles. Then, select the candidate (Candidate A or Candidate B) whom you would be more likely to invite for an interview for that specific role. There are no 'right' or 'wrong' answers; we are interested in your professional judgment. Please make your choices as if you were genuinely hiring for this role.

Please find below the job description for the role of Office Assistant:

Position: Office Assistant

Job Type: Full-time, Permanent

Overview: We are seeking a reliable, proactive and organized Office Assistant to drive efficiency in our day-to-day administrative operations. This role requires exceptional attention to detail, excellent communication skills, and the ability to manage multiple tasks with foresight.

Key Responsibilities:

- Manage calendars, proactively schedule meetings to optimize time for all participants, and coordinate appointments.
- Handle incoming communications (phone, email, mail), employing systems to prioritize and pre-draft responses.
- Maintain organized filing systems (physical and digital) designed for rapid information retrieval.
- Prepare documents, synthesize information for reports, and draft professional correspondence.
- Assist with data entry and ensure data integrity through validation and cleanup.
- Support office supply management and implement predictive inventory tracking.
- Provide administrative support to team members.
- Coordinate with vendors and service providers as needed.

Required Qualifications:

- High school diploma or equivalent 2+ years of administrative or clerical experience.
- Proficiency in Microsoft Office Suite (Word, Excel, Outlook).
- Strong written and verbal communication skills.
- Excellent organizational and time management abilities.
- Professional demeanor and customer service orientation.
- Ability to work independently and as part of a team.

Preferred Qualifications:

- Associate's degree or higher in Business Administration or related field.
- Knowledge of basic bookkeeping or accounting principles.
- Experience with modern productivity and office management software that enhances efficiency.
- Familiarity with basic data analysis or business intelligence concepts.

### Graphic designer, 6-month contract

You will now be shown pairs of candidate profiles for the 6 months role of Graphic Designer at an undefined organization and industry. For each pair, please carefully review both candidate profiles. Then, select the candidate (Candidate A or Candidate B) whom you would be more likely to invite for an interview for that specific role. There are no 'right' or 'wrong' answers; we are interested in your professional judgment. Please make your choices as if you were genuinely hiring for this role.

Please find below the job description for the role of Graphic Designer:

Position: Graphic Designer

Job Type: Full-time, 6-month contract

Overview: We are seeking a creative, strategic, efficient, and detail-oriented Graphic Designer to elevate our brand's visual identity. This role requires a strong conceptual thinker with excellent design skills, who can translate marketing and product objectives into compelling visual narratives that engage and convert our target audience.

Key Responsibilities:

- Develop visual concepts and execute original designs for a wide range of digital and print assets, including marketing campaigns, social media graphics, web banners, presentations, and branding collateral.
- Collaborate with marketing, product, and sales teams to ensure brand consistency and create cohesive visual experiences across all touchpoints.
- Maintain and evolve the company's brand identity guidelines, ensuring all creative output is aligned with our visual standards.
- Manage design projects from brief to final delivery, effectively managing timelines and stakeholder feedback.
- Prepare and preflight files for both print and digital production, ensuring technical accuracy and high-quality output.
- Manage and organize a library of design assets, templates, and stock photography for efficient team-wide access.
- Stay current with design trends, industry best practices, and emerging technologies to continuously elevate our creative work.
- Support the creative team with various design tasks as needed.

Required Qualifications:

- Associate's degree or equivalent 3+ years of professional graphic design experience.
- A strong portfolio showcasing a range of creative projects and a high level of design skill.
- Expert proficiency in Adobe Creative Suite (Photoshop, Illustrator, InDesign).
- Excellent understanding of typography, color theory, layout, and visual hierarchy.
- Strong communication and presentation skills, with the ability to articulate design concepts.
- Proven ability to manage multiple projects simultaneously and meet deadlines.

Preferred Qualifications:

- Bachelor's degree in a subject related to graphic design.
- Experience with UI/UX design principles and proficiency in tools like Figma or Sketch.
- Basic knowledge of motion graphics (e.g., After Effects) or video editing.
- Experience with modern design systems and workflow tools that enhance creative efficiency.
- Familiarity with basic user experience (UX) research or A/B testing concepts as they relate to design effectiveness.

### Graphic designer, permanent contract

You will now be shown pairs of candidate profiles for the permanent role of Graphic Designer at an undefined organization and industry. For each pair, please carefully review both candidate profiles. Then, select the candidate (Candidate A or Candidate B) whom you would be more likely to invite for an interview for that specific role. There are no 'right' or 'wrong' answers; we are interested in your professional judgment. Please make your choices as if you were genuinely hiring for this role.

Please find below the job description for the role of Graphic Designer:

Position: Graphic Designer

Job Type: Full-time, permanent contract

Overview: We are seeking a creative, strategic, efficient, and detail-oriented Graphic Designer to elevate our brand's visual identity. This role requires a strong conceptual thinker with excellent design skills, who can translate marketing and product objectives into compelling visual narratives that engage and convert our target audience.

Key Responsibilities:

- Develop visual concepts and execute original designs for a wide range of digital and print assets, including marketing campaigns, social media graphics, web banners, presentations, and branding collateral.
- Collaborate with marketing, product, and sales teams to ensure brand consistency and create cohesive visual experiences across all touchpoints.
- Maintain and evolve the company's brand identity guidelines, ensuring all creative output is aligned with our visual standards.
- Manage design projects from brief to final delivery, effectively managing timelines and stakeholder feedback.
- Prepare and preflight files for both print and digital production, ensuring technical accuracy and high-quality output.
- Manage and organize a library of design assets, templates, and stock photography for efficient team-wide access.

- Stay current with design trends, industry best practices, and emerging technologies to continuously elevate our creative work.
- Support the creative team with various design tasks as needed.

Required Qualifications:

- Associate's degree or equivalent 3+ years of professional graphic design experience.
- A strong portfolio showcasing a range of creative projects and a high level of design skill.
- Expert proficiency in Adobe Creative Suite (Photoshop, Illustrator, InDesign).
- Excellent understanding of typography, color theory, layout, and visual hierarchy.
- Strong communication and presentation skills, with the ability to articulate design concepts.
- Proven ability to manage multiple projects simultaneously and meet deadlines.

Preferred Qualifications:

- Bachelor's degree in a subject related to graphic design.
- Experience with UI/UX design principles and proficiency in tools like Figma or Sketch.
- Basic knowledge of motion graphics (e.g., After Effects) or video editing.
- Experience with modern design systems and workflow tools that enhance creative efficiency.
- Familiarity with basic user experience (UX) research or A/B testing concepts as they relate to design effectiveness.

### Software engineer, 6-month contract

You will now be shown pairs of candidate profiles for the 6-month role of Software Engineer at an undefined organization and industry. For each pair, please carefully review both candidate profiles. Then, select the candidate (Candidate A or Candidate B) whom you would be more likely to invite for an interview for that specific role. There are no 'right' or 'wrong' answers; we are interested in your professional judgment. Please make your choices as if you were genuinely hiring for this role.

Please find below the job description for the role of Software Engineer:

Position: Software Engineer

Job Type: Full-time, 6 month

Overview: We are seeking an innovative, analytical, and results-driven Software Engineer to join our dynamic technical team. This role requires a strong problem-solver with a passion for building robust, scalable, and high-quality software. The ideal candidate will be a collaborative team player who can translate complex requirements into clean, efficient, and maintainable code.

Key Responsibilities:

- Design, develop, test, and deploy high-performance software applications and systems that meet user and business needs.
- Collaborate with product managers, designers, and other engineers to define technical requirements and implement new features.
- Write clean, well-documented, and testable code, adhering to best practices and established coding standards.
- Maintain and improve existing software systems, including troubleshooting bugs, enhancing performance, and reducing technical debt.

- Participate actively in code reviews to ensure code quality, share knowledge, and foster a collaborative development environment.
- Create and maintain comprehensive technical documentation for systems, APIs, and processes to ensure clarity and continuity.
- Contribute to the continuous improvement of our technology stack and development processes.

Required Qualifications:

- Associate's degree in Computer Science or a related field, or an equivalent 3+ years of professional software engineering experience.
- A portfolio of projects (e.g., via GitHub) showcasing proficiency in software development.
- Solid proficiency in at least one modern programming language (e.g., Python, Java, C++, JavaScript).
- Strong understanding of fundamental computer science concepts, including data structures, algorithms, and object-oriented design.
- Experience with version control systems, particularly Git.
- Excellent problem-solving, debugging, and analytical skills.

Preferred Qualifications:

- Bachelor's degree in Computer Science, Software Engineering, or a related technical field.
- Demonstrated ability to quickly learn and apply new and emerging technologies to solve problems and enhance efficiency.
- Experience with cloud computing platforms (e.g., AWS, Azure, GCP).
- Familiarity with Agile development methodologies and CI/CD pipelines.
- Experience with database technologies, including both SQL and NoSQL databases.

### Software engineer, permanent contract

You will now be shown pairs of candidate profiles for the permanent role of Software Engineer at an undefined organization and industry. For each pair, please carefully review both candidate profiles. Then, select the candidate (Candidate A or Candidate B) whom you would be more likely to invite for an interview for that specific role. There are no 'right' or 'wrong' answers; we are interested in your professional judgment. Please make your choices as if you were genuinely hiring for this role.

Please find below the job description for the role of Software Engineer:

Position: Software Engineer

Job Type: Full-time, permanent contract

Overview: We are seeking an innovative, analytical, and results-driven Software Engineer to join our dynamic technical team. This role requires a strong problem-solver with a passion for building robust, scalable, and high-quality software. The ideal candidate will be a collaborative team player who can translate complex requirements into clean, efficient, and maintainable code.

Key Responsibilities:

- Design, develop, test, and deploy high-performance software applications and systems that meet user and business needs.

- Collaborate with product managers, designers, and other engineers to define technical requirements and implement new features.
- Write clean, well-documented, and testable code, adhering to best practices and established coding standards.
- Maintain and improve existing software systems, including troubleshooting bugs, enhancing performance, and reducing technical debt.
- Participate actively in code reviews to ensure code quality, share knowledge, and foster a collaborative development environment.
- Create and maintain comprehensive technical documentation for systems, APIs, and processes to ensure clarity and continuity.
- Contribute to the continuous improvement of our technology stack and development processes.

Required Qualifications:

- Associate's degree in Computer Science or a related field, or an equivalent 3+ years of professional software engineering experience.
- A portfolio of projects (e.g., via GitHub) showcasing proficiency in software development.
- Solid proficiency in at least one modern programming language (e.g., Python, Java, C++, JavaScript).
- Strong understanding of fundamental computer science concepts, including data structures, algorithms, and object-oriented design.
- Experience with version control systems, particularly Git.
- Excellent problem-solving, debugging, and analytical skills.

Preferred Qualifications:

- Bachelor's degree in Computer Science, Software Engineering, or a related technical field.
- Demonstrated ability to quickly learn and apply new and emerging technologies to solve problems and enhance efficiency.
- Experience with cloud computing platforms (e.g., AWS, Azure, GCP).
- Familiarity with Agile development methodologies and CI/CD pipelines.
- Experience with database technologies, including both SQL and NoSQL databases.

## Annex C - Sentiment Analysis

This section documents the procedure used to classify recruiter sentiment toward artificial intelligence (AI) based on free-text responses to survey item C1 (“Please describe the key factors that influenced your decisions…”).

### *Data Preparation and Labeling Procedure*

We analyze sentiment only in responses that explicitly reference AI. Mentions were identified using basic string matching (e.g., “AI”, “ai”, “a.i.”, “artificial intelligence”). Of all respondents who answered question C1, 34 percent contained at least one reference to AI and were retained for analysis.

Sentiment labeling was conducted using GPT-5.1 via the OpenAI API. Each text was classified into one of four mutually exclusive categories: *positive*, *neutral*, *negative*, or *missing* (the latter indicating insufficient information). The model was prompted using the following instruction:

```
You will receive a list of short texts from a survey among HR recruiters. For each
text, determine the sentiment towards AI skills ONLY, using one of these labels:

- "positive": recruiter views the candidate’s AI skills as an advantage.
- "neutral": AI skills are mentioned factually without a clear sentiment.
- "negative": recruiter views the candidate’s AI skills as a disadvantage or
concern.
- "missing": Sentiment cannot be determined.

Examples:
Positive: "qualified to use AI tools" or "AI skills" or "AI usage" → positive
Neutral: "How they discussed the use of AI." → neutral
Negative: "Candidates who can work independent of AI." or "I preferred
candidates who did not use AI" → negative

IMPORTANT: Output ONLY the label word: positive, neutral, negative, or missing.

text: {text}
```

A manual audit of the labeled data revealed 52 responses that GPT-5.1 had classified as *neutral* but which, given their minimal phrasing and context (e.g., “AI skills”, “AI usage”), should be considered *positive*. These labels were corrected accordingly.

### *Overall Sentiment Distribution*

Among responses that mentioned AI, sentiment is predominantly positive:

Table C.1: Overall AI sentiment distribution

| Sentiment | Percent |
| --- | --- |
| Negative | 30.7 |
| Positive | 63.5 |
| Neutral | 2.5 |
| Missing | 3.4 |

### *Heterogeneity Across Recruiter Subgroups*

This analysis provides a descriptive overview of the heterogeneity in recruiters' evaluations of AI skills across demographic and occupational subgroups.

#### *Occupation*

Recruiters in graphic design express much more negative sentiment toward AI compared to those in office administration or software engineering. A chi square test shows that these differences are highly statistically significant ($\chi^2 = 26.83$, df = 4, p = 2.15e-05, $p < 0.001$).

Table C.2: Sentiment toward AI by role (%)

| Role | Negative | Positive |
| --- | --- | --- |
| Graphic Designer | 43.8 | 49.0 |
| Office Assistant | 24.8 | 70.1 |
| Software Developer | 22.0 | 72.9 |

#### *AI Usage in Own Work*

Self-reported AI usage (question A8_1) was recoded into *Low* (never to monthly use) and *High* (all higher frequencies). Recruiters with low AI usage exhibit substantially more negative sentiment.

Table C.3: Sentiment toward AI by AI usage of recruiters (%)

| Recruiter AI usage | Negative | Positive |
| --- | --- | --- |
| Low | 55.4 | 41.8 |
| High | 19.4 | 73.4 |

#### *Age*

Younger recruiters tend to be more negative toward AI than older recruiters.

Table C.4: Sentiment toward AI by age group (%)

| Age group (in years) | Negative | Positive |
|---|---|---|
| 18–29 | 43.1 | 49.2 |
| 30-44 | 35.2 | 58.8 |
| 45+ | 22.0 | 72.8 |

*Gender*

Female recruiters show more negative sentiment toward AI than male recruiters.

Table C.5: Sentiment toward AI by gender (%)

| Gender | Negative | Positive |
|---|---|---|
| Female | 34.2 | 61.1 |
| Male | 25.0 | 67.3 |

*Additional Characteristics*

Contract type (permanent vs. short-term) and country of residence (United Kingdom vs. United States) do not show meaningful differences in sentiment toward AI.

***Correlation between Recruiter AI usage and Sentiment toward AI***

Across all roles, recruiters who use AI more heavily tend to have more positive sentiment toward it, with this pattern being strongest for the Office Assistant role and weakest (but still present) for the Software Developer role. Self-reported AI usage (question A8_1) was recoded into *Low* (never to monthly use) and *High* (all higher frequencies).
Combined with the insights from Table C.2, it appears that recruiters are more skeptical about AI usage in Graphic Design. This skepticism appears to reflect genuine concerns about AI in creative work rather than being driven by lower AI usage among these recruiters. AI usage among recruiters for the different roles is at comparable levels[3] while the sentiment toward AI is significantly lower when hiring for Graphic Design. The relationship between high AI usage and positive sentiment toward AI is positive and statistically significant regardless of which role a recruiter is hiring for.

Table C.6: Correlation between recruiter AI usage and recruiter sentiment toward AI

| Role | High AI usage | Positive toward AI | N | Phi correlation | Chi Square |
|---|---|---|---|---|---|
| Graphic Designer | 62.6 | 51.6 | 182 | 0.298 | 16.2*** |

[3] A chi square test on AI usage x role on the full dataset reveals that the two variables are independent with a test-statistic of 5.4 and p-value = 0.067. This makes intuitive sense as recruiters were not invited to the survey for hiring for a specific role but were randomly assigned one of the three job advertisements.

| | | | | | |
|---|---|---|---|---|---|
| Office Assistant | 71.7 | 72.1 | 247 | 0.389 | 37.4*** |
| Software Developer | 69.0 | 74.1 | 116 | 0.242 | 6.8*** |

Note: ***p<0.01

Table C.7 shows some examples of free-text responses about what factors influenced choices from respondents by occupation they were shown.

Table C.7: Sample of free-text responses by occupation, variable C1

| **Role** | **Positive sentiment example** | **Neutral sentiment example** | **Negative sentiment example** |
|---|---|---|---|
| Graphic Designer | 'The main factor guiding my choices was how closely each candidate's skills and experience aligned with the role requirements. I placed the most value on proven ability with Adobe Creative Suite, branding and UX/UI design, since those are central to the position. Education and certifications were also important when they directly supported the role, while innovative use of AI were considered an advantage but secondary to strong practical design and branding expertise.' | 'Candidates who can work independent of AI' | 'I refrained from selecting anyone that listed leveraging or using AI in their application. AI is a powerful tool but I would much rather hire a graphic designer that can do their job without needing to rely on AI.' |
| Office Assistant | 'I chose candidates with experience in AI automation and calendar management over the candidates that did not have that type of experience' | 'AI experience' | 'I don't like people who rely on AI as they typically have reduced critical thinking skills after a while so tried to avoid that as much as I could.' |
| Software Developer | 'Skills section - in | 'AI competency' | 'The ones that had actual |
| | addition to programming the use of agile methodologies and Gen AI’ | | degrees and not AI degrees’ |

## Annex D - Recruiters' background

This section details the demographic, professional, and attitudinal profiles of the study participants. The final sample consists of 1,725 unique recruiters and HR professionals. The sample is predominantly based in the United Kingdom and the United States, with a high level of professional experience and varied adoption levels of Artificial Intelligence in their workflows.

### *Respondent Demographics*

The sample is diverse in terms of gender, education, and ethnicity. The majority of respondents identified as Female (60.2%) and White (72.8%). The sample is highly educated, with over 69% holding a Bachelor's degree or higher.

Table D.1 - Demographic Characteristics of Respondents (N = 1,725)

| **Characteristic** | **Category** | **Count (N)** | **Percentage (%)** |
|---|---|---|---|
| Country | United Kingdom | 921 | 53.4% |
| | United States | 729 | 42.3% |
| | Germany | 58 | 3.4% |
| | Missing / No response | 17 | 1.0% |
| Gender | Female | 1,039 | 60.2% |
| | Male | 657 | 38.1% |
| | Non-binary / Other | 9 | 0.5% |
| | Prefer not to say | 3 | 0.2% |
| | Missing / No response | 17 | 1.0% |
| Age | Mean | 43.4 years | |
| Ethnicity | White | 1,255 | 72.8% |
| | Black or African American | 209 | 12.1% |
| | Asian | 105 | 6.1% |
| | Mixed or Multiple ethnicities | 61 | 3.5% |
| | Hispanic or Latino | 49 | 2.8% |
| | Other | 15 | 0.9% |
| | Prefer not to say | 14 | 0.8% |
| | Missing / No response | 17 | 1.0% |
| Education Level | Bachelor's or equivalent | 760 | 44.1% |
| | Master's or equivalent | 443 | 25.7% |
| | Post-secondary non-tertiary | 161 | 9.3% |
| | Upper secondary or lower | 159 | 9.2% |
| | Short-cycle tertiary | 120 | 7.0% |
| | Doctoral or equivalent | 53 | 3.1% |
| | Prefer not to say | 12 | 0.7% |
| | Missing / No Response | 17 | 1.0% |

### *Professional Profile*

Participants represent a wide range of industries, with the largest segments being Health Care (12.9%) and Professional Services (12.5%). The majority work in large organizations (over 250 employees) and identify primarily as HR Generalists rather than domain specialists.

Table D.2 - Professional background (N = 1,725)

| **Characteristic** | **Category** | **Count (N)** | **Percentage (%)** |
|---|---|---|---|
| Current Professional Role | HR | 772 | 44.8% |
| | Recruiter / Recruitment Consultant | 293 | 17.0% |
| | Talent Acquisition | 142 | 8.2% |
| | CV / Professional Development Coach | 71 | 4.1% |
| | Other | 278 | 16.1% |
| | Prefer not to say | 152 | 8.8% |
| | Missing / No Response | 17 | 1.0% |
| Professional Experience | Mean Years of Experience | 10.56 | |
| | Median Years of Experience | 7.00 | |
| Organization Size | Large (> 250 employees) | 891 | 51.7% |
| | Medium (50-250 employees) | 434 | 25.2% |
| | Small (< 50 employees) | 346 | 20.1% |
| | Prefer not to say / Don't know | 37 | 2.1% |
| | Missing / No Response | 17 | 1.0% |
| Specialization | Generalist | 1,067 | 61.9% |
| | Domain Specialist | 581 | 33.7% |
| | Missing / No Response | 77 | 4.5% |
| Industry Sector | Health Care and Social Assistance | 223 | 12.9% |
| | Professional, Scientific, and Technical Services | 216 | 12.5% |
| | Educational Services | 158 | 9.2% |
| | Finance and Insurance | 134 | 7.8% |
| | Public Administration | 119 | 6.9% |
| | Retail Trade | 117 | 6.8% |
| | Manufacturing | 114 | 6.6% |
| | Other Services (except Public Administration) | 108 | 6.3% |
| | Information | 96 | 5.6% |
| | Construction | 59 | 3.4% |
| | Transportation and Warehousing | 54 | 3.1% |
| | Accommodation and Food Services | 50 | 2.9% |
| | Prefer not to say | 50 | 2.9% |
| | Arts, Entertainment, and Recreation | 49 | 2.8% |
| | Management of Companies and Enterprises | 38 | 2.2% |
| | Real Estate and Rental and Leasing | 27 | 1.6% |
| | Administrative and Support and Waste Management | 24 | 1.4% |
| | Agriculture, Forestry, Fishing and Hunting | 23 | 1.3% |
| | Utilities | 17 | 1.0% |
| | Wholesale Trade | 17 | 1.0% |
| | Don't know | 10 | 0.6% |
| | Mining, Quarrying, and Oil and Gas Extraction | 5 | 0.3% |
| | Missing / No Response | 17 | 1.0% |

### *Technology Adoption & Attitudes*

The survey measured the adoption of Applicant Tracking Systems (ATS) and Artificial Intelligence tools. While general AI usage (e.g., ChatGPT) is high, with 67.0% using it weekly or daily in their general work, specific usage for HR purposes is lower, with 36.3% using it weekly or daily.

Table D.3 - Technology Adoption and Opinions (N = 1,725)

| **Survey Question** | **Response Category** | **Count (N)** | **Percentage (%)** |
|---|---|---|---|
| Frequency of AI Usage (general work) | 5 - Daily | 521 | 30.2% |
| | 4 - Weekly | 634 | 36.8% |
| | 3 - Monthly | 192 | 11.1% |
| | 2 - Less than once a month | 232 | 13.4% |
| | 1 - Never | 129 | 7.5% |
| | Missing | 17 | 1.0% |
| Frequency of AI Usage (HR purposes) | 5 - Daily | 214 | 12.4% |
| | 4 - Weekly | 412 | 23.9% |
| | 3 - Monthly | 315 | 18.3% |
| | 2 - Less than once a month | 315 | 18.3% |
| | 1 - Never | 452 | 26.2% |
| | Missing | 17 | 1.0% |
| Prevalence of ATS in hiring | 5 - Used for all screening tasks | 209 | 12.1% |
| | 4 - Frequently used | 509 | 29.5% |
| | 3 - Occasionally used | 414 | 24.0% |
| | 2 - Rarely used | 225 | 13.0% |
| | 1 - Not used at all | 351 | 20.3% |
| | Missing | 17 | 1.0% |
| Perceived AI Impact on industry | Extremely impactful | 237 | 13.7% |
| | Very impactful | 644 | 37.3% |
| | Moderately impactful | 503 | 29.2% |
| | Slightly impactful | 242 | 14.0% |
| | Not impactful at all | 82 | 4.8% |
| | Missing | 17 | 1.0% |

### *Valuation of candidate signals*

Recruiters were asked to rate the value of different candidate credentials on a 5-point scale. Traditional signals such as Years of Experience and Formal Education remain the most highly valued, with 85.8% and 62.8% of recruiters rating them as "Very" or "Extremely" valuable, respectively.

Among AI-specific signals, relevant AI skills were rated highest, while micro-credentials (regardless of provider) were generally viewed as having moderate to slight value.

Table D.4 - Recruiter valuation of hiring signals. Question: "How valuable are the following signals when evaluating a candidate?"

| **Hiring Signal** | **Not Valuable at all** | **Slightly Valuable** | **Moderately Valuable** | **Very Valuable** | **Extremely Valuable** |
|---|---|---|---|---|---|
| Years of relevant work experience | 0.5% | 2.4% | 11.3% | 38.2% | 47.6% |
| Level of formal education | 1.2% | 8.6% | 27.5% | 40.2% | 22.6% |
| Relevant AI skills | 6.0% | 18.8% | 23.9% | 32.8% | 18.6% |
| University Degree specifically in AI | 7.2% | 15.2% | 26.1% | 31.9% | 19.7% |
| University-branded micro-credentials | 6.6% | 20.3% | 32.6% | 30.9% | 9.6% |
| Company-branded micro-credentials | 7.2% | 22.5% | 34.4% | 26.7% | 9.2% |
| Micro-credentials (generic platforms) | 6.8% | 23.0% | 35.8% | 26.1% | 8.3% |

## Annex E - Details on Survey Flow and Data Processing

### *Survey Flow*

The survey followed a structured flow. First, participants were randomly assigned to one of the three roles (Office Assistant, Graphic Designer, or Software Engineer). Within each role, they were further randomized to either the 6-month or permanent contract group. Before the comparison tasks, participants were shown a detailed job description for the specific role and contract type they were assigned to. The main part of the survey consisted of 15 comparison tasks, where participants selected which candidate they would be more likely to invite for an interview. After the comparison tasks, participants answered a series of control questions about their background, professional experience, demographic characteristics, and familiarity and attitude toward AI, including a question about their frequency of generative AI usage. Open-text questions probed their decision-making process in the CV comparison tasks.

### *Data Processing*

The raw data collected from Qualtrics underwent a systematic data processing pipeline to prepare it for analysis. This process, automated with Python scripts to ensure replicability, transformed the survey output into an analysis-ready dataset. The core procedure involved mapping each recruiter's choice in the conjoint tasks to the full set of characteristics corresponding to the two CVs they evaluated, creating a structured dataset linking the hiring outcome to the specific experimental conditions, such as the type of AI-skill credential presented, and the attributes of each candidate, including their designated advantages or disadvantages.

Subsequently, the data was transformed from a wide format, where each row represented a respondent, into a long (or "tidy") format. This reshaping resulted in a final dataset where each row represents a single candidate evaluation (i.e., one candidate being either selected for an interview or not), which is the standard structure required for estimating choice models.

## Annex F - Example CVs and Choice task

### Choice task

Which one of the following two candidates would you invite for an interview if you were hiring for a 6-month role as Office Assistant? Please consult the candidate profiles below. Please only select one candidate.

- ☐ Candidate A
- ☐ Candidate B

### Candidate A

**Age:** 32
**Gender:** Female
**Professional Summary**
Profile with 9 years of administrative experience and a BA in Communication Studies.

**Education**
**National Louis University**
*BA in Communication Studies*, 2011-2015
**Triton College**
*Associate of Arts*, 2009-2011

**Work Experience**
**Administrative Assistant**, Horizon Enterprises (2020-Present)

- Delivered key administrative support for the finance and operations departments.

**Office Clerk**, Lake County Health Dept (2015-2020)

- Handled a variety of clerical tasks, including data processing, document management, and official correspondence.

**Skills**

- Microsoft Office Suite
- Google Workspace
- Calendar Management
- Data Entry
- Agile Methodologies

**Candidate B**
**Age:** 61
**Gender:** Female

**Professional Summary**

Profile with 9 years of administrative experience and a BA in Public Administration. Uses AI tools to enhance productivity and automate routine tasks.

**Education**

**Roosevelt University**

*BA in Public Administration*, 1986-1990

**McHenry County College**

*Associate of Arts*, 1984-1986

**Work Experience**

**Administrative Assistant**, Crestview Inc. (2020-Present)

- Provided administrative support, utilizing AI to automate scheduling and draft official communications.

**Office Clerk**, Cook County Health Dept (2015-2020)

- Carried out clerical duties with a focus on data integrity and maintaining efficient records.

**Skills**

- Microsoft Office Suite
- Google Workspace
- AI-Powered Scheduling
- Automated Reporting
- Prompt Engineering

## Annex G - Codebook

The codebook detailed the variables contained in the dataset, with the variables divided into two main sections:

1. **Respondent-level variables:** Demographic, professional, and attitudinal data for each participant.
2. **Experimental task variables:** Variables related to the conjoint choice experiment, detailing the tasks presented and the attributes of the candidates shown.

### *Respondent Demographics*

Personal demographic information provided by the respondent.

| Column Name | Question / Description | Format / Answers |
|---|---|---|
| Q53 | "By clicking 'I consent', you confirm..." | Categorical: "I consent" |
| A1 | "Please enter your age in years." | Numeric |
| A2 | "Please enter your gender." | Categorical: "Male", "Female", "Non-binary / third gender", "Prefer not to say" |
| A3 | "Which of the following industry do you primarily recruit for?" | Categorical: Dropdown list of 22 industries. |
| A5 | "In which of the following countries are you based in?" | Categorical: Dropdown list of 190 countries. |
| A9 | "What is the highest level of education you have achieved?" | Categorical: 7 levels (e.g., "Upper secondary...", "Bachelor's...", "Doctoral...") |
| A12 | "What is your ethnicity?" | Categorical: 7 levels (e.g., "White", "Black or African American", "Asian"...) |

### *Respondent Profile (HR/Recruiting)*

Information on the respondent's professional role and background.

| Column Name | Question / Description | Format / Answers |
|---|---|---|
| Q120 | "Are you a Prolific participant?" | Categorical: "Yes", "No" |
| Q118 | "What is your Prolific participant ID?" | String (Open-text) |
| A4 | "Please enter your years of work experience in recruiting or HR." | Numeric |
| A6 | "What is the size of your organization?" | Categorical: 5 levels (e.g., "Small (< 50 employees)", "Medium (50-250)...") |
| A7 | "What is your role?" | Categorical: 6 levels (e.g., "Recruiter...", "HR", "Talent Acquisition"...) |
| Q123 | "Do you consider yourself a domain specialist (sector-related) or a generalist?" | Categorical: "Domain specialist", "Generalist" |
| Q126 | "If you selected 'Other', please specify your role below:" | String (Open-text, follow-up to A7) |

### *Respondent Opinions & Behaviors*

Attitudinal and behavioural questions related to AI and recruiting practices.

| Column Name | Question / Description | Format / Answers |
|---|---|---|
| A8_1 | "How often do you use AI tools (e.g., ChatGPT, Claude) in your own work?" | Ordinal (1-5 Scale): 1="Never", 5="Daily" |
| Q76_1 | "How often do you use AI tools (e.g., AI-powered sourcing platforms, automated candidate scoring in your ATS) for HR purposes?" | Ordinal (1-5 Scale): 1="Never", 5="Daily" |
| A10_1 | "How prevalent are Applicant Tracking Systems (ATS) in candidate screening in your organization?" | Ordinal (1-5 Scale): 1="Not used at all", 5="Used for all..." |
| A11_1 | "According to your perception, how impactful is AI going to be in the industry you are primarily recruiting for?" | Ordinal (1-5 Scale): 1="Not impactful at all", 5="Extremely impactful" |
| C1 | "Please describe the key factors that influenced your decisions..." | String (Open-text) |
| C2_1 | "How valuable are... **Relevant AI skills**" | Ordinal (1-5 Scale): 1="Not valuable at all", 5="Extremely valuable" |
| C2_2 | "How valuable are... **Micro-credentials (e.g., from Coursera, edX)**..." | Ordinal (1-5 Scale): 1="Not valuable at all", 5="Extremely valuable" |
| C2_3 | "How valuable are... **University-branded online micro-credentials**..." | Ordinal (1-5 Scale): 1="Not valuable at all", 5="Extremely valuable" |
| C2_4 | "How valuable are... **Company-branded micro-credentials**..." | Ordinal (1-5 Scale): 1="Not valuable at all", 5="Extremely valuable" |
| C2_5 | "How valuable are... **University Degree specifically in AI**..." | Ordinal (1-5 Scale): 1="Not valuable at all", 5="Extremely valuable" |
| C2_6 | "How valuable are... **Years of relevant work experience**" | Ordinal (1-5 Scale): 1="Not valuable at all", 5="Extremely valuable" |
| C2_7 | "How valuable are... **Level of formal education**" | Ordinal (1-5 Scale): 1="Not valuable at all", 5="Extremely valuable" |

### *Experimental Task Variables*

This section details the columns related to the conjoint choice tasks. These columns are structurally generated based on a combination of **Role**, **Contract Type**, and **Task ID**.

#### *Naming Convention*

All experimental columns follow this naming convention:
*[ROLE]_[CONTRACT]_[TASK_ID]_[DESCRIPTOR]*.

***[ROLE]*:**

- **OA**: Office Assistant
- **GD**: Graphic Designer
- **SE**: Software Engineer

***[CONTRACT]*:**

- **6M**: 6-Month Contract
- **PE**: Permanent

***[TASK_ID]*:**

- **A1 - A19**, **B1 - B19**: These 38 unique IDs refer to the scenarios defined in the CV combinations.xlsx file. The A cases and disadvantage cases represent different experimental conditions (e.g., disadvantage cases include a counter-certificate for the advantaged candidate).
- **Example 1 (A10):** Disadvantage: 'Older', AI Treatment: 'None'
- **Example 2 (B14):** Disadvantage: 'Older', AI Treatment: 'AI: LinkedIn'

***[DESCRIPTOR]***

For each unique combination of [ROLE], [CONTRACT], and [TASK_ID], the following 9 columns are generated.

| # | [DESCRIPTOR] Suffix | Column Name Example | Description | Format / Answers |
|---|---|---|---|---|
| 1 | (none) | GD_6M_A10 | The respondent's choice for this task. | Categorical: "Candidate A", "Candidate B" |
| 2 | _Treated_Candidate | GD_6M_A no 10_Treated_Candidate | Which candidate (A or B) received the AI treatment. | Categorical: "Candidate A", "Candidate B" |
| 3 | _Treatment_Type | GD_6M_A10_Treatment_Type | The type of AI skill/certificate shown. | Categorical: "None", "Self", "LinkedIn", "University", "Company" |
| 4 | _CandidateA_Age | GD_6M_A10_CandidateA_Age | Age attribute of Candidate A. | Categorical: "Young", "Old" |
| 5 | _CandidateB_Age | GD_6M_A10_CandidateB_Age | Age attribute of Candidate B. | Categorical: "Young", "Old" |
| 6 | _CandidateA_Gender | GD_6M_A10_CandidateA_Gender | Gender attribute of Candidate A. | Categorical: "Male", "Female" |
| 7 | _CandidateB_Gender | GD_6M_A10_CandidateB_Gender | Gender attribute of Candidate B. | Categorical: "Male", "Female" |
| 8 | _CandidateA_Education | GD_6M_A10_CandidateA_Education | Education attribute of Candidate A. | Categorical: "Higher Education", "Lower Education" |
| 9 | _CandidateB_Education | GD_6M_A10_CandidateB_Education | Education attribute of Candidate B. | Categorical: "Higher Education", "Lower Education" |

# Annex H - Additional Results

**Table H1.** Invitation probability of AI-skilled candidates in paired CV evaluations. Using CV pairs as the unit of observation, the dependent variable indicates whether the AI candidate is invited. Models control for job and recruiter characteristics and compare no disadvantage, age disadvantage, and educational disadvantage, with recruiter AI usage showing the strongest positive association with selecting the AI-skilled candidate.

| | *Dependent variable: AI candidate gets invite (0/1)* | | | | | | | | | | | |
|---|---|---|---|---|---|---|---|---|---|---|---|---|
| | Role Type | | | + Recruiter Characteristics | | | + Recruiter AI Impact | | | + Recruiter AI Usage | | |
| | No Dis. | Age | Education | No Dis. | Age | Education | No Dis. | Age | Education | No Dis. | Age | Education |
| **Role (ref. Graphic Designer)** | | | | | | | | | | | | |
| Office Assistant | 0.04 | 0.51*** | 0.57*** | 0.004 | 0.50*** | 0.58*** | −0.09 | 0.47*** | 0.53*** | −0.08 | 0.47*** | 0.53*** |
| | (0.09) | (0.07) | (0.08) | (0.10) | (0.07) | (0.08) | (0.10) | (0.07) | (0.08) | (0.10) | (0.07) | (0.08) |
| Software Engineer | 0.36*** | 0.43*** | 0.34*** | 0.34*** | 0.45*** | 0.35*** | 0.28*** | 0.42*** | 0.33*** | 0.27*** | 0.42*** | 0.32*** |
| | (0.10) | (0.07) | (0.08) | (0.10) | (0.07) | (0.08) | (0.10) | (0.07) | (0.08) | (0.10) | (0.07) | (0.08) |
| **Contract Type (ref. 6 months)** | | | | | | | | | | | | |
| Permanent | 0.09 | −0.17*** | −0.07 | 0.06 | −0.19*** | −0.08 | 0.02 | −0.20*** | −0.10 | 0.01 | −0.20*** | −0.10* |
| | (0.08) | (0.06) | (0.06) | (0.08) | (0.06) | (0.06) | (0.08) | (0.06) | (0.06) | (0.08) | (0.06) | (0.06) |
| **Recruiter** | | | | | | | | | | | | |
| Age | | | | 0.02*** | 0.0001 | 0.0002 | 0.02*** | 0.0003 | 0.0004 | 0.02*** | 0.0003 | 0.0004 |
| | | | | (0.003) | (0.001) | (0.001) | (0.004) | (0.001) | (0.001) | (0.004) | (0.001) | (0.001) |
| Male (ref. Female) | | | | 0.12 | −0.08 | 0.18*** | 0.08 | −0.12* | 0.14** | 0.05 | −0.12* | 0.12* |
| | | | | (0.08) | (0.06) | (0.06) | (0.08) | (0.06) | (0.06) | (0.08) | (0.06) | (0.06) |
| **Recruiting for ICT (ref. Other)** | | | | | | | | | | | | |
| | | | | 1.07*** | −0.12 | 0.20 | 0.92*** | −0.21 | 0.08 | 0.87*** | −0.21 | 0.04 |
| | | | | (0.23) | (0.13) | (0.14) | (0.23) | (0.13) | (0.14) | (0.23) | (0.13) | (0.14) |
| **AI impact on labour market (ref. No impact)** | | | | | | | | | | | | |
| Slightly impactful | | | | | | | 0.67*** | 0.16 | 0.37** | 0.57*** | 0.13 | 0.31 |
| | | | | | | | (0.21) | (0.18) | (0.19) | (0.21) | (0.18) | (0.19) |
| Moderately impactful | | | | | | | 1.10*** | 0.62*** | 0.89*** | 0.92*** | 0.59*** | 0.77*** |
| | | | | | | | (0.20) | (0.17) | (0.18) | (0.20) | (0.17) | (0.18) |
| Very impactful | | | | | | | 1.43*** | 0.73*** | 1.02*** | 1.18*** | 0.70*** | 0.83*** |
| | | | | | | | (0.20) | (0.16) | (0.17) | (0.21) | (0.17) | (0.18) |
| Extremely impactful | | | | | | | 1.67*** | 0.89*** | 1.26*** | 1.39*** | 0.87*** | 1.02*** |
| | | | | | | | (0.22) | (0.18) | (0.19) | (0.23) | (0.19) | (0.20) |
| **AI Usage (ref. Never)** | | | | | | | | | | | | |
| Less than once a month | | | | | | | | | | 0.35* | 0.21 | 0.14 |
| | | | | | | | | | | (0.18) | (0.14) | (0.16) |
| Monthly | | | | | | | | | | 0.82*** | 0.21 | 0.33** |
| | | | | | | | | | | (0.20) | (0.15) | (0.16) |
| Weekly | | | | | | | | | | 0.60*** | 0.12 | 0.34** |
| | | | | | | | | | | (0.17) | (0.13) | (0.15) |
| Daily | | | | | | | | | | 0.70*** | 0.13 | 0.51*** |
| | | | | | | | | | | (0.18) | (0.14) | (0.15) |
| Constant | 0.24*** | −0.51*** | −0.52*** | −0.38 | −1.83*** | −0.92 | −1.03 | −2.31*** | −1.60*** | −1.22* | −2.34*** | −1.58*** |
| | (0.08) | (0.06) | (0.06) | (0.65) | (0.64) | (0.58) | (0.69) | (0.66) | (0.60) | (0.68) | (0.65) | (0.60) |
| Observations | 2,765 | 4,720 | 4,442 | 2,765 | 4,700 | 4,428 | 2,765 | 4,700 | 4,428 | 2,765 | 4,700 | 4,428 |
| Log Likelihood | −1,849.64 | −3,193.07 | −3,014.44 | −1,820.35 | −3,170.09 | −2,997.83 | −1,765.14 | −3,129.96 | −2,944.17 | −1,752.70 | −3,128.03 | −2,934.85 |
| Akaike Inf. Crit. | 3,707.27 | 6,394.14 | 6,036.88 | 3,666.70 | 6,366.19 | 6,021.67 | 3,572.28 | 6,301.91 | 5,930.34 | 3,555.39 | 6,306.06 | 5,919.70 |

*Note:* *p<0.1; **p<0.05; ***p<0.01

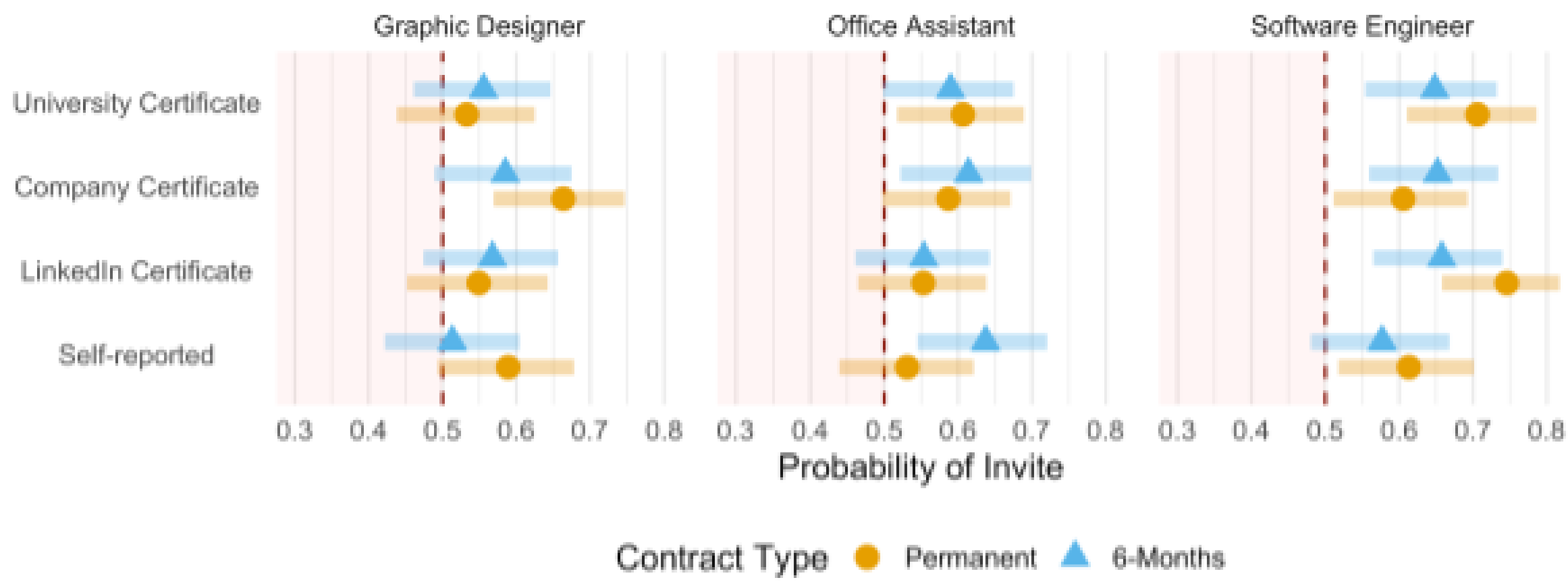

**Figure A1**: Contract type (permanent vs. 6-months) does not substantially affect the relationship between AI skills and interview invitations. Points show predicted probabilities of receiving an interview invitation for each contract type, estimated from a logistic regression. Bars represent 95% confidence intervals. The dashed horizontal line at 0.5 marks the baseline of equal likelihood of invitation for either of the two applicants.

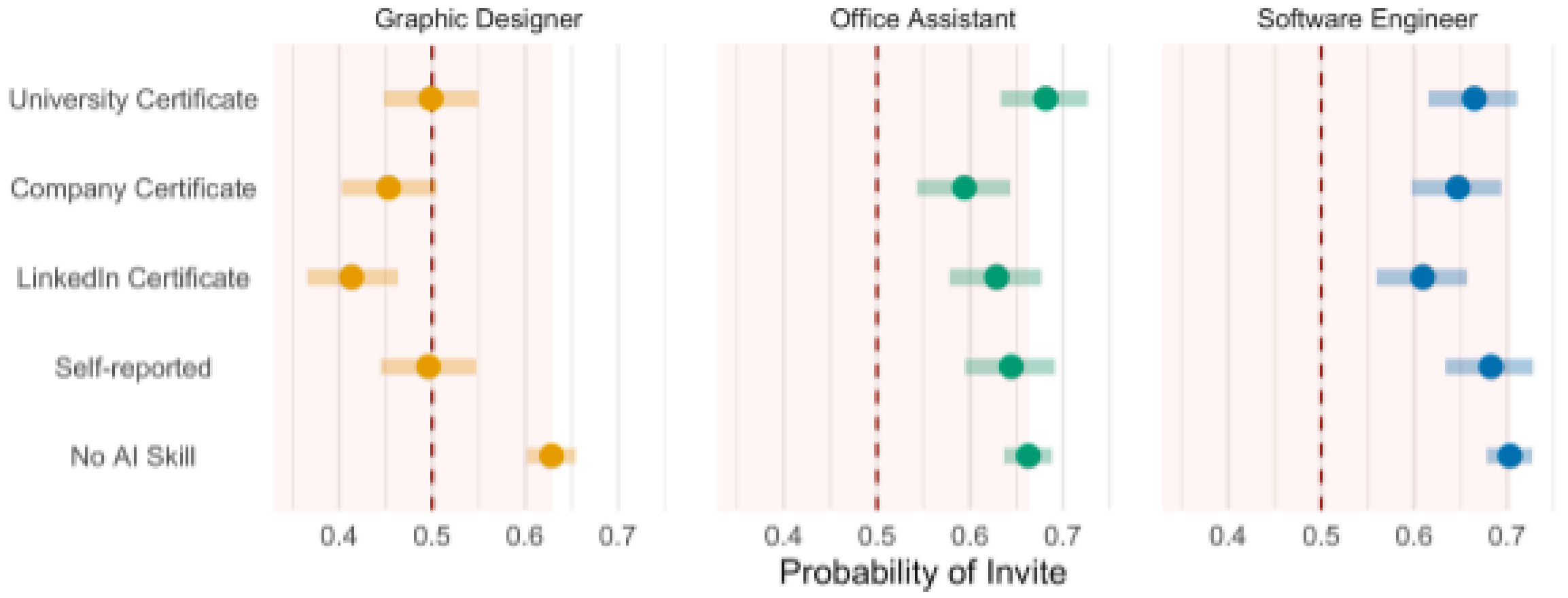

**Figure A2**: CV with gender disadvantage (female with and without AI skills vs. male without AI skills). Points show predicted probabilities of receiving an interview invitation for each AI skill level, estimated from a logistic regression. Bars represent 95% confidence intervals. The shaded red area indicates the baseline probability of being invited for an interview for CVs without AI skills. The dashed horizontal line at 0.5 marks equal likelihood of invitation for either of the two applicants. We can see that women have a higher likelihood of being invited than men. For the Graphic Design role having AI skills reduces the likelihood of being invited for an interview. For the Office Assistant and the Software Engineer role AI skills do not significantly change the likelihood of being invited for an interview compared to the baseline.

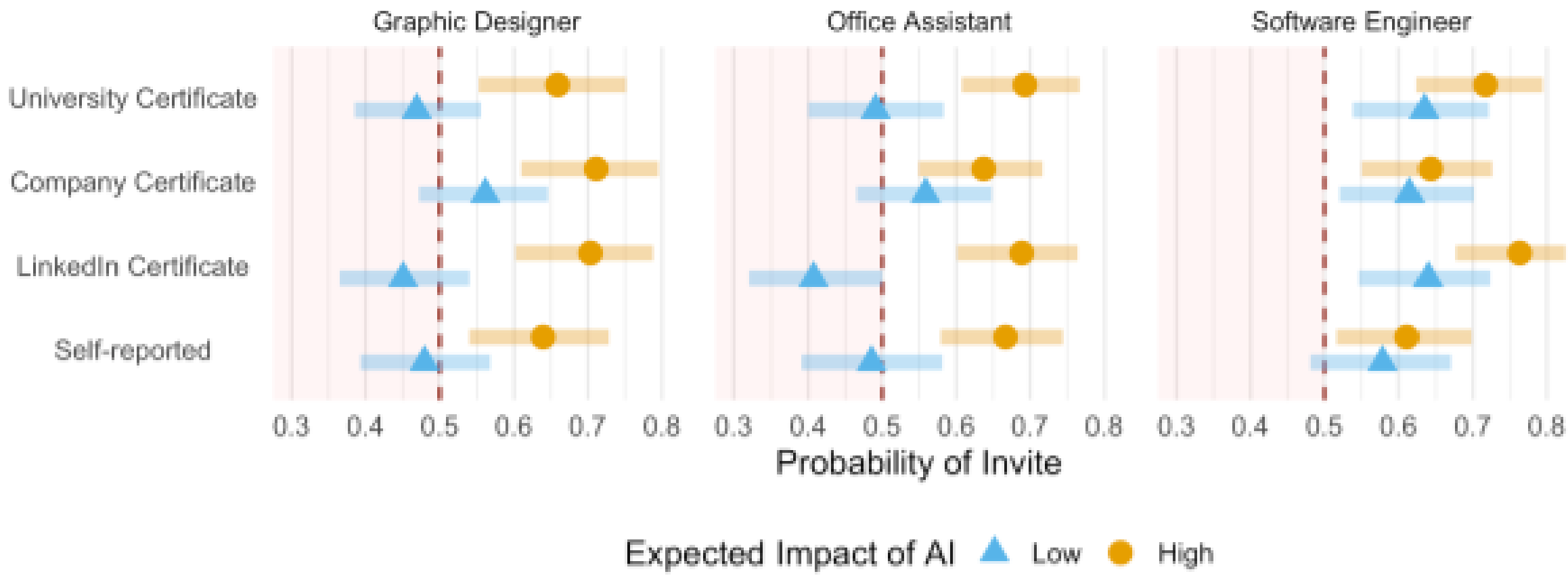

**Figure A3**: The valuation of AI skills is heavily dependent on the recruiter's own expectations about the future importance of AI revealing a pronounced gatekeeper effect. Recruiters who view AI as highly impactful (very or extremely impactful) strongly reward AI-skilled candidates, whereas recruiters who perceive AI as having low impact (not impactful to moderately impactful) show a flattened response. Points show predicted probabilities of receiving an interview invitation for each AI skill level, estimated from a logistic regression. Bars represent 95% confidence intervals. The dashed horizontal line at 0.5 marks the baseline of equal likelihood of invitation for either of the two applicants.

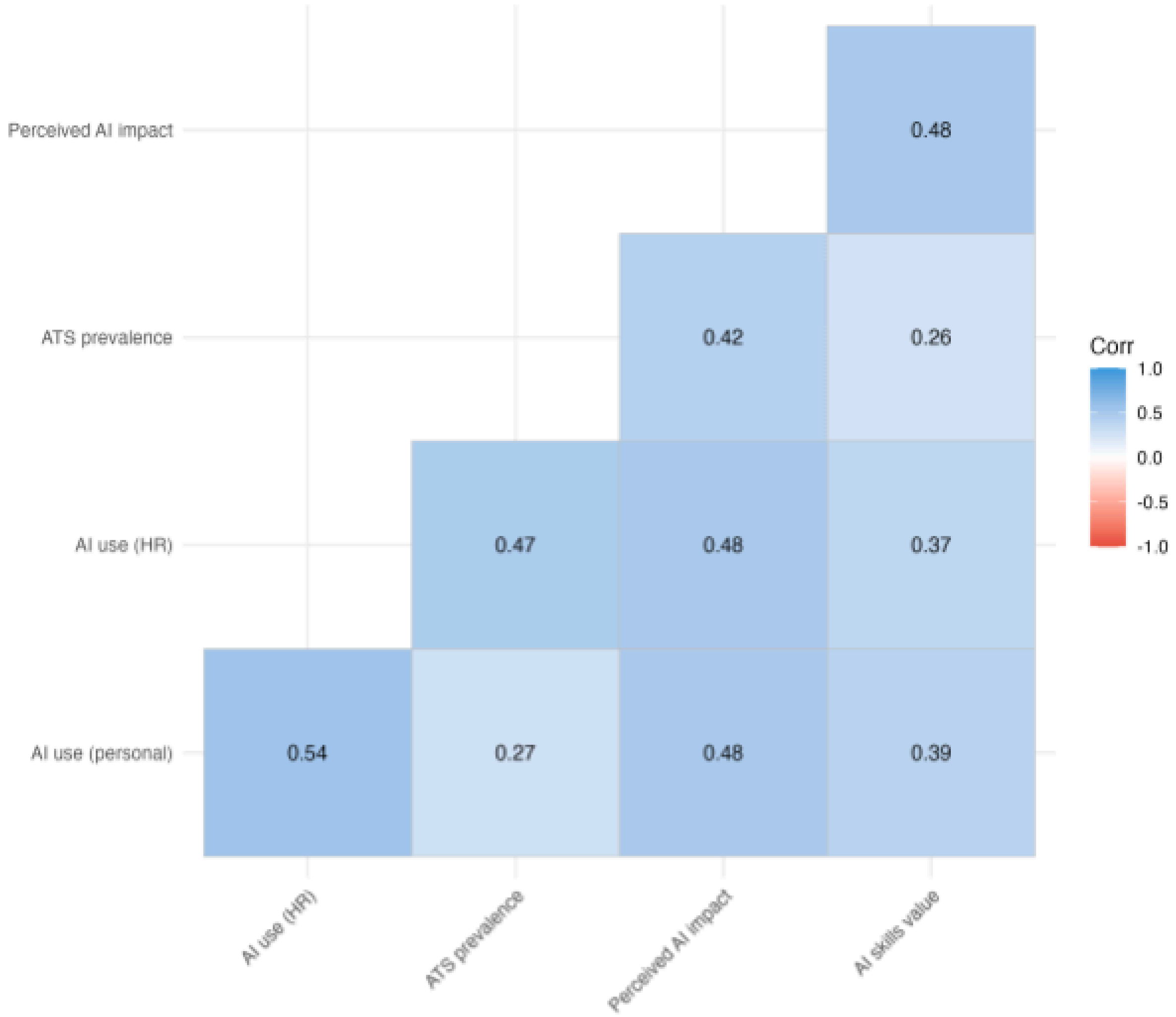

**Figure A4**: Spearman rank correlations among AI adoption variables (N = 1,556 recruiters, only complete cases without missing values). Results are consistent with polychoric correlation estimates reported in Figure A4.

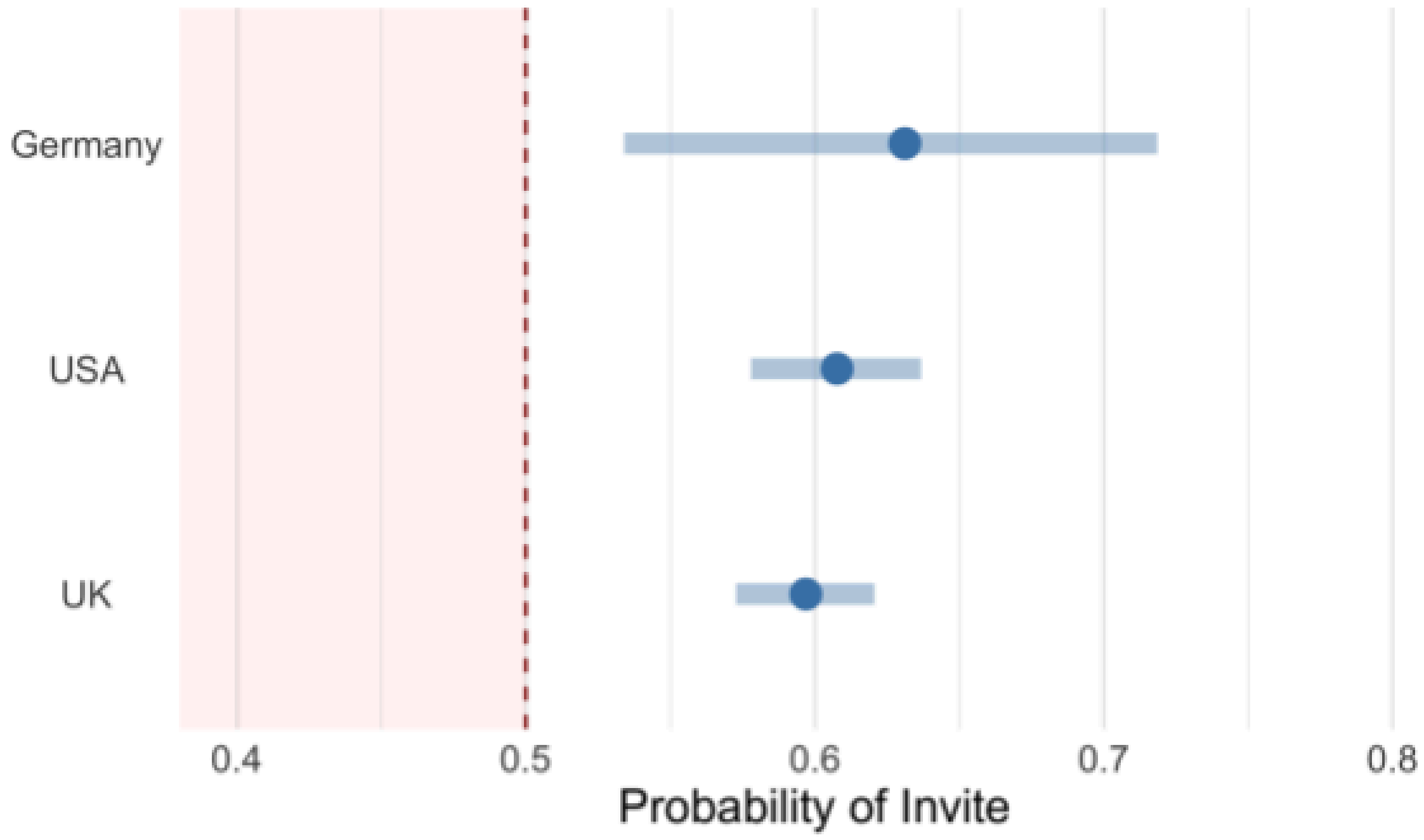

**Figure A5**: Country-level comparison. For Germany the sample size is significantly smaller with only 57 respondents. Points show predicted probabilities of receiving an interview invitation for all AI skill levels combined, estimated from a logistic regression. Bars represent 95% confidence intervals. The dashed horizontal line at 0.5 marks the baseline of equal likelihood of invitation for either of the two applicants.

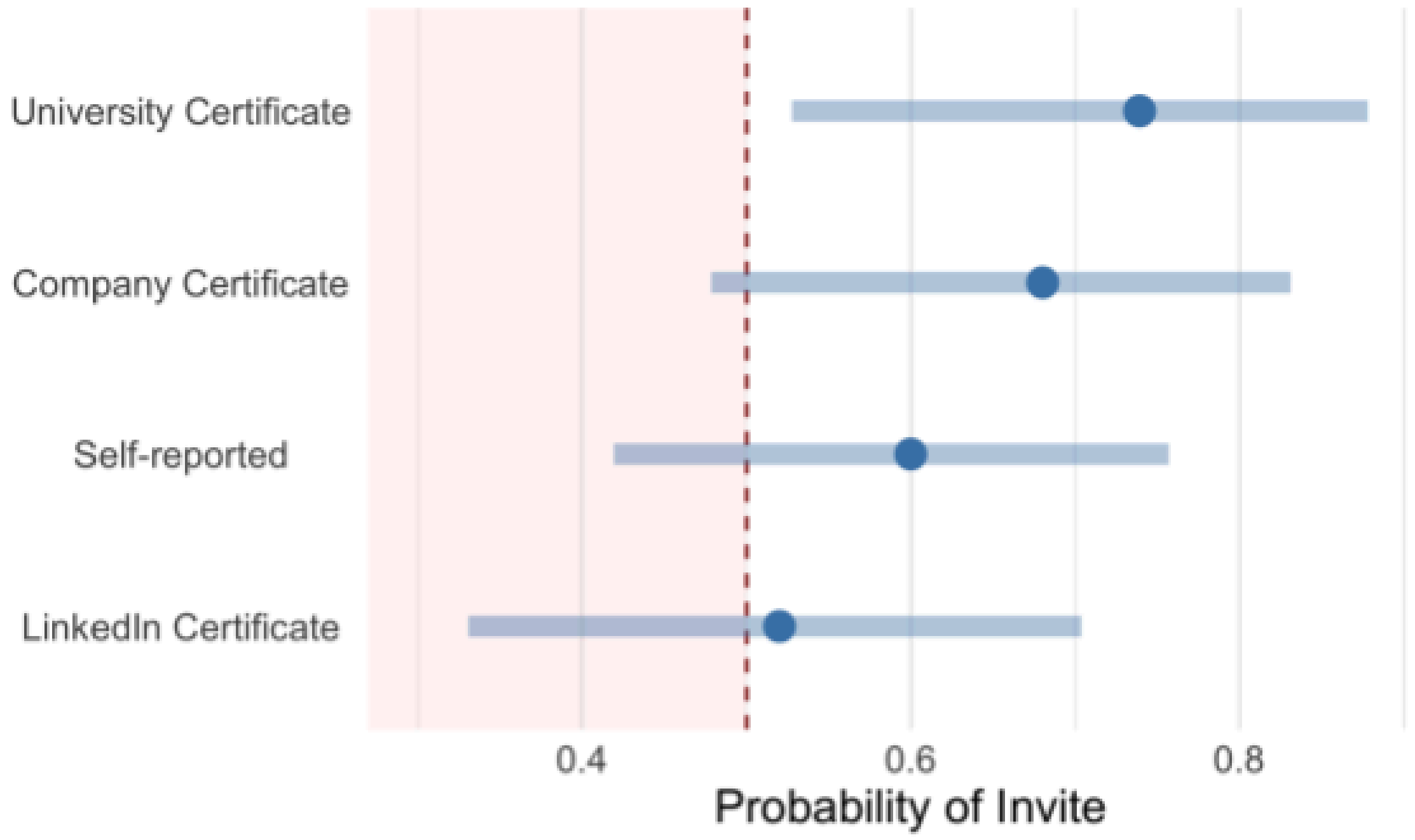

**Figure A6**: Value of AI skills among German recruiters. For Germany the sample size is significantly smaller with only 57 respondents compared to the UK and U.S.. Points show predicted probabilities of receiving an interview invitation for each AI skill level, estimated from a logistic regression. Bars represent 95% confidence intervals. The dashed horizontal line at 0.5 marks the baseline of equal likelihood of invitation for either of the two applicants.

**Table A1**: AI usage by recruiter demographics.

| Dimension | Category | High AI Usage (%) | Low AI usage (%) | N |
|---|---|---|---|---|
| Gender | Female | 64.9 | 35.1 | 974 |
| | Male | 73.4 | 26.6 | 557 |
| Age | Under 30 | 65.9 | 34.1 | 246 |
| | 30-39 | 69.1 | 30.9 | 528 |
| | 40-49 | 69.6 | 30.4 | 372 |
| | 50+ | 66.2 | 33.8 | 385 |
| Country | UK | 65.6 | 34.4 | 867 |
| | US | 71.1 | 28.9 | 664 |

**Table A2**: Expectations about future importance of AI usage by recruiter demographics.

| Dimension | Category | High Impact (%) | Low Impact (%) | N |
|---|---|---|---|---|
| Gender | Female | 49.4 | 50.6 | 974 |
| | Male | 57.8 | 42.2 | 557 |
| Age | Under 30 | 53.3 | 46.7 | 246 |
| | 30-39 | 50.6 | 49.4 | 528 |
| | 40-49 | 57.3 | 42.7 | 372 |
| | 50+ | 49.9 | 50.1 | 385 |
| Country | UK | 49.6 | 50.4 | 867 |
| | US | 56.2 | 43.8 | 664 |